\documentclass[superscriptaddress, prd, aps,amsmath,amssymb,showpacs,showkeys, onecolumn]{revtex4-2}
\usepackage[dvips]{graphicx,color}
\usepackage{subfigure}
\usepackage{times}
\usepackage{xcolor}
\usepackage{tikz}
\usepackage[%
  colorlinks=true,
  urlcolor=blue,
  linkcolor=red,
  citecolor=blue
]{hyperref}
\usepackage{orcidlink}

\begin{document}
\title{Impact of Thermodynamic Corrections on the Stability of Hayward-Anti de Sitter Black Hole Surrounded by a Fluid of Strings}

\author{Shyamalee Bora\orcidlink{0009-0009-2605-9530} }
\email[Email: ]{swamaleebora@gmail.com}

\affiliation{Department of Physics, Tezpur University, Tezpur, 784028, Assam, India.}

\author{Dhruba Jyoti Gogoi\orcidlink{0000-0002-4776-8506}}
\email[Email: ]{moloydhruba@yahoo.in}

\affiliation{Department of Physics, Moran College, Moranhat, Charaideo 785670, Assam, India.}
\affiliation{Research Center of Astrophysics and Cosmology, Khazar University, 41 Mehseti Street, AZ1096 Baku, Azerbaijan.}

\author{Pralay Kumar Karmakar \orcidlink{0000-0002-3078-9247}}
\email[Email: ]{pkk@tezu.ernet.in}
\affiliation{Department of Physics, Tezpur University, Tezpur, 784028, Assam, India.}

\begin{abstract}

We explore the modified thermodynamics of a Hayward-Anti de Sitter (H-AdS) black hole in atypical conditions, incorporating a string fluid, Hayward regularisation, and quantum entropy corrections. Our analysis reveals a first-order phase transition between small and large black hole phases, characterised by a swallowtail behaviour in the Gibbs free energy profiles. It is found that the key parameters - string fluid strength, Hayward regularisation scale, and quantum correction coefficients significantly influence the critical temperature and phase stability of the H-AdS system. It is further noticed that a large black hole phase is stabilised by these modifications, with quantum corrections smoothing the transition. This model offers a valuable framework to study quantum gravity effects on black hole thermodynamics with potential implications in analysing black hole evolution and astrophysical observations.

\end{abstract}
	

\maketitle
\section{Introduction}\label{sec01}
The discovery of black holes as thermal objects revolutionised a new era in modern physics that attempted to integrate theories of general relativity, quantum mechanics, and thermodynamics. The solutions to the equations of general relativity allowed for the description of black holes. This started with K. Schwarzschild in 1916, when he provided the first solution to Einstein's field equations to model a spherically symmetric, non-rotating black hole \cite{Schwarzschild:1916uq}. Some other observations, such as the generation of gravitational waves, show perfect agreement of black hole physics with the predictions of general relativity \cite{LIGOScientific:2016aoc, LIGOScientific:2017ycc, Ghosh:2019mnu}. 
The fact that black holes can be studied as a thermodynamic system gained a great deal of attention in the 1970s \cite{Bekenstein:1973ur,Bekenstein:1972tm}. In 1971, S. Hawking demonstrated that the area of the event horizon of a black hole always increases \cite{Hawking:1971tu}. It shows an analogy between the area of a black hole's event horizon and the second law of thermodynamics. A year later, J. D. Bekenstein proposed the groundbreaking area law \cite{Bekenstein:1973ur}. It stated that the entropy of a black hole is directly proportional to the area of its event horizon. This confirmed, for the first time, that a thermodynamic parameter can be introduced in the black hole description. S. Hawking further strengthened the thermodynamic interpretation of black holes \cite{Hawking:1974rv, Hawking:1975vcx}. He put forward the concept of Hawking radiation, according to which the black hole's event horizon emits thermal radiation due to quantum effects close to it. The temperature is inversely proportional to its mass \cite{Hawking:1974rv, Hawking:1975vcx}. This type of radiative phenomenon could be described by applying quantum field theory in curved spacetime.

In the above context, assuming that Einstein's field equation is correct, it was observed that its black hole solutions follow a set of laws similar to those of thermodynamics. Here, temperature ($T$) and entropy ($S$) are represented by the surface gravity ($\kappa$) and the area of the black hole horizon ($A$), respectively \cite{Bekenstein:1973ur}. In the works of Bardeen {\it et al.} \cite{Bardeen:1973gs}, the four laws of black hole thermodynamics were formulated. According to the zeroth law, the surface gravity is constant across the event horizon. The first law states that energy conservation includes changes in mass, angular momentum, and magnetic charge. The second law states that the total horizon area never decreases. According to the third law, the reduction of the surface gravity to zero is impossible. In such advancements, black holes and conventional thermodynamic systems were thoroughly compared. They also brought to light certain unanswered questions. Hawking radiation, in contrast to the classical predictions, implied that black holes can eventually evaporate due to a gradual loss of mass and energy. This black hole information paradox \cite{Mathur:2009hf} is a problem that classical general relativity was unable to resolve on its own. It suggested the necessity of a quantum theory of gravity, which means that a proper explanation of black hole thermodynamics requires the application of quantum gravity.

Many researchers have rigorously studied the thermodynamics of black holes \cite{Carlip:2014pma, Al-Badawi:2025rcq, Wald:2025nbz, Ahmed:2025iqz, Chunaksorn:2025nsl, Witten:2024upt, Elizalde:2025iku, Yang:2024krx, Bakopoulos:2024zke, Ong:2022frf, Bouzenada:2025ypa, Sen:2024kdr, Ahissou:2025nhq}. String theory and quantum gravity studies indicate that microscopic degrees of freedom are the sources of black hole entropy \cite{Krasnov:2009pd, Domagala:2004jt, Carlip:2007za, Oriti:2018qty}. Loop quantum gravity connects entropy to quantum states of the horizon and offers similar discoveries \cite{Wald:1993nt, Dong:2025qnp}. Quantum fluctuations disrupt the thermodynamics of black holes as their size decreases \cite{-ul-islam:2023ivu, Hamil:2023dmx}. This leads to a modification of the entropy associated with a black hole. Several attempts have been made in recent years to discuss leading-order corrections to black hole thermodynamical parameters, such as those arising from quantum effects, loop quantum gravity, string theory, and other extensions of general relativity, so that a deeper insight into the microscopic structure of spacetime can be provided \cite{Jacobson:1995ab, Fursaev:1994te,         Gogoi:2024ypn}. Logarithmic corrections to black hole entropy are predicted by quantum gravity methods such as loop quantum gravity and string theory \cite{Maziashvili:2015uwa, Abreu:2020dyu, Asin:2014gta}. These corrections result from the finite size of the phase space related to black hole microstates and quantum fluctuations of the horizon \cite{Kaul:2000kf}. Higher-order components in the entropy, including inverse powers
of the area, are predicted by quantum field theory in curved spacetime and manifest in theories with higher curvature terms, such as effective field theories or f(R) gravity. In 1997, J. Maldacena proposed a key concept in string theory named the AdS/CFT correspondence \cite{Solodukhin:2011gn}. The CFT's temperature, entropy, and correlation functions are directly mapped to the thermodynamic characteristics of black holes, including temperature, entropy, and their corrections. Therefore, any modification to black hole thermodynamics will alter the dual CFT \cite{Maldacena:1997re}. For testing the AdS/CFT correspondence, more accurate predictions in holographic dualities are made possible by corrected black hole thermodynamics \cite{Chen:2023pgs}. Astrophysicists investigate quantum gravity effects and how they appear in dual field theories by examining corrections in AdS space \cite{Ladghami:2024wkv, Abdusattar:2023xxs}. Understanding the function of black holes in the larger holographic principle and the quantum structure of spacetime requires these modifications and refinements in the basic formalism.

Pourhassan {\it et al.} \cite{Pourhassan:2022opb} previously used quantum gravitationally corrected entropy to analyse the thermodynamics of a Schwarzschild-Tangherlini AdS black hole of quantum size. The authors have also used Jarzynski equality to produce quantum work between black hole states. Also, non-perturbative corrections to Gibbs free energy and specific heat were examined in that work. It was explored how partition weights and quantum work relate to evaporating AdS black holes. In Ref. \cite{Dehghani:2021qzm}, the authors observed that the classical Bekenstein-Hawking entropy is modified by quantum phenomena, bringing logarithmic corrections to the black hole entropy.  The quantum corrections in the Born-Infeld and logarithmic models result in the second-order phase transition, which shows that the black holes' thermodynamic stability has changed. It is also concluded that behaviours of the system can only be accurately described by a modification of the first law of black hole thermodynamics in the presence of logarithmic corrections. Another study \cite{Pourhassan:2022cvn} examines the effects of quantum corrections on the thermodynamics of Born-Infeld BTZ black holes in massive gravity by incorporating both exponential and logarithmic corrections to the black hole entropy. A comparative analysis, conducted between the corrected and classical thermodynamic quantities, reveals that exponential corrections strongly influence the Helmholtz free energy of larger black holes and lead to the appearance of a second point in the first-order phase transition, altering the standard phase structure. The authors have also derived the equation of state for the exponentially corrected black hole, resulting in a leading-order virial expansion. These findings demonstrate how quantum corrections, particularly in the framework of massive gravity, significantly impact the thermodynamic behaviours and phase transitions of black holes. 

R. Ali {\it et al.} \cite{Ali:2022yar} also looked at how thermal fluctuations affect the corrected entropy. When the thermodynamic behaviour of black holes is studied in Horndeski gravity by incorporating quantum corrections via the Generalised Uncertainty Principle (GUP), the results show that adding logarithmic corrections improves the black hole system's thermodynamic stability, as shown by the heat capacity studies. The results reduce to classical predictions when the corrective parameter is removed. An application of the Kaniadakis statistics alters the thermodynamic parameters, such as temperature, entropy, and heat capacity, for rotating AdS black holes \cite{Hazarika:2024lnx}. This analysis also reveals changes in the phase transition behaviour of these black holes. A similar study of the thermodynamic behaviour of rotating AdS black holes through the Kaniadakis thermo-statistical lens finds notable deviations from predictions based on the traditional Boltzmann-Gibbs statistics \cite{Gohain:2024eer}. S. Mohapatra {\it et al.} \cite{Mahapatra:2011si} used fluctuation moments, represented in terms of black hole response coefficients, to construct the leading-order entropy corrections. An analysis of mass and charge fluctuations in regular charged (R-charged) AdS black holes leads to the discovery of universality in the logarithmic corrections to entropy across various spacetime dimensions. It suggests that such corrections are intrinsic features of black hole thermodynamics, independent of specific black hole characteristics. 
The authors in Ref. \cite{Mahapatra:2011si} derive the leading-order entropy corrections using fluctuation moments, which are expressed in terms of black hole response coefficients. They analyse mass and charge fluctuations in R-charged AdS black holes and discover universality in the logarithmic corrections to entropy across various spacetime dimensions. This universality suggests that such corrections are intrinsic features of black hole thermodynamics, independent of specific black hole characteristics.
Apart from these studies, there are several other studies which deal with different aspects related to corrected black hole thermodynamics as found in the literature \cite{Hamil:2023zeb, Ndongmo:2023qac, Tan:2024jkj, Song:2023ebj, Xiao:2023two, Cassani:2024tvk, Gogoi:2024scc, Mandal:2023ahb, Ladghami:2024qaf, Sudhanshu:2024wqb}.

Motivated by the discussion above, we will explore the higher-order entropy corrections for a Hayward-AdS (H-AdS) black hole surrounded by a fluid of strings. Additionally, we will examine various thermal properties of this black hole system, deriving corrected expressions for its enthalpy, Helmholtz free energy, thermodynamic volume, specific heat, internal energy, and Gibbs free energy. Furthermore, we will investigate how the presence of the string fluid alters the black hole's geometry and thermodynamics, thereby offering fresh insights into its stability and entropy.

Throughout this paper, we stick to the metric signature $(+,-,-,-)$ and consider geometrical units (fundamental constants) $c=G=\hbar=k_{B}=1$. After this brief introduction in Section \ref{sec01}, the rest of the paper is arranged as follows. In Section \ref{sec2}, we briefly overview the geometry of Hayward-AdS black holes surrounded by a fluid of strings. In Section \ref{section3}, we discuss some basic thermodynamic properties followed by the thermodynamic fluctuations, and the corrected thermodynamics is discussed in Section \ref{section4}. In Section \ref{section5}, phase transitions, critical temperature, and stability of black hole systems are discussed. Finally, we summarise our investigated results in Section \ref{sec7} and in Section \ref{sec8} we conclude our discussion.

\section{Hayward-AdS black hole surrounded by a fluid of strings} \label{sec2}
The spacetime for the Hayward black hole can be obtained by solving Einstein's equation paired with a nonlinear electromagnetic field, which serves as the black hole's physical source. In Ref. \cite{Nascimento:2023tgw}, Nascimento {\it et al.} successfully generated a set of spherically symmetric black hole solutions including the presence of two other sources, i.e., cosmological constant and fluid of strings. It generalises the original Hayward black hole picture. This system can be described by an action given as
\begin{equation}
S= \frac{1}{16\pi}\int d^4x\sqrt{-g}(R+\mathcal{L})\label{eq_action}.
\end{equation}
Here, $g$ denotes the determinant of the metric tensor, $g_{\mu\nu}$, $R$ is the scalar curvature and $\mathcal{L}$ gives the Lagrangian density of the nonlinear electromagnetic field \cite{Hayward:2005gi, Waseem:2025bwb, Nascimento:2023tgw}.

Due to the presence of coupling between gravitational and non-linear electromagnetic fields, the action is modified to yield the equation as follows
\begin{equation}
G_{\mu\nu} = 2 \left(\frac{\partial\mathcal{L}(F)}{\partial F}F_{\mu\sigma}F^{\sigma}_{\,\,\nu}-\frac{1}{4}g_{\mu\nu}\mathcal{L}(F)\right),
\label{equação de einstein mista}
\end{equation}
where, $G_{\mu\nu}=R_{\mu\nu}-\frac{1}{2}Rg_{\mu\nu}$.

Since the source of Hayward's solution arises from nonlinear electrodynamics \cite{Ayon-Beato:2000mjt}, the Lagrangian density can be represented as
\begin{equation}
\mathcal{L}(F)=\frac{6(2l^2F)^{3/2}}{\kappa^2l^2[1+(2l^2F)^{3/4}]^2}.
\label{Hayward's Lagrangian}
\end{equation}
Here, $\mathcal{L}(F)$ represents the source of the Hayward black hole solution in the context of the nonlinear electrodynamics framework and $l$ is the Hayward parameter whose value is restricted in the range $0\leq l \leq \infty$. It should also be noted that the Hayward parameter $l$, as considered in Hayward's original work, is of the order of Planck's length \cite{Nascimento:2023tgw, Bronnikov:2000vy}.  In Eq.(\ref{Hayward's Lagrangian}), the electromagnetic scalar $F=F^{\mu \nu}F_{\mu \nu}$ is represented in a nonlinear form, where $F_{\mu \nu}$ is the Maxwell-Faraday tensor \cite{Waseem:2025bwb, Nascimento:2023tgw}.

Using Eq.(\ref{Hayward's Lagrangian}), the components of the energy-momentum tensor are obtained as \cite{Ayon-Beato:2000mjt, Nascimento:2023tgw}
\begin{equation}
    T_{t}^{\;t}=T_{r}^{\;r}=\frac{12l^2m^2}{(r^3+2l^2m)^{2}} ,
    \label{eq4}
\end{equation}
\begin{equation}
 T_{\theta}^{\;\theta}=T_{\phi}^{\;\phi}=-\frac{24(r^3-l^2m)l^2m^2}{(r^3+2l^2m)^{3}}.
 \label{eq5}
\end{equation}
 In the above equations, $(l,m)$ are constants and these are taken to be positive \cite{Nascimento:2023tgw}. 

The terms $-\Lambda g_{\mu\nu}$, $T_{\mu\nu}^{\text{FS}}$ are added to the right-hand side and left-hand side, respectively, of Eq. (\ref{equação de einstein mista}), so that the cosmological constant and fluid of strings are taken into account. This leads to
\begin{equation}
R_{\mu\nu}-\frac{1}{2}Rg_{\mu\nu}-\Lambda g_{\mu\nu}= 2 \left(\frac{\partial\mathcal{L}(F)}{\partial F}F_{\mu\sigma}F^{\sigma}_{\,\,\nu}-\frac{1}{4}g_{\mu\nu}\mathcal{L}(F)\right)+ T_{\mu\nu}^{\text{FS}},
\label{geral_eq}
\end{equation}
here, $-\Lambda g_{\mu\nu}$ corresponds to the cosmological constant, and $T_{\mu\nu}^{\text{FS}}$ is the energy-momentum tensor of the fluid of strings. With the first term referring to the Einstein equation modified by the coupling with a nonlinear electromagnetic field and the second term referring to the fluid of strings, the right-hand side of the equation above represents an effective energy-momentum tensor.
As assumed by Nascimento {\it et al.} \cite{Nascimento:2023tgw} the stress-energy tensor for the fluid of strings has the components.
\begin{equation}
T_{t}^{\;t}= T_{r}^{\;r}= -\frac{\varepsilon}{r^2}\left(\frac{b}{r}\right)^{2/\beta},
\label{eq:1.17.1}
\end{equation}
\begin{equation}
T_{\theta}^{\;\theta}= T_{\phi}^{\;\phi}= \frac{\varepsilon}{\beta r^2}\left(\frac{b}{r}\right)^{2/\beta},
\label{eq:1.18.1}
\end{equation}
where $b$ is a positive constant of integration. This parameter controls the strength of the string fluid. In other words, it gives the intensity of the fluid of the strings. It has the physical dimension of length and is kept positive so that unphysical behaviour like imaginary contributions or ill-defined expressions having logarithms or powers can be prevented \cite{Waseem:2025bwb}. $\varepsilon=\pm1$ determines the sign of the energy density of the fluid of strings. $\beta$ is a dimensionless constant controlling the effect of the power-law decay of the string fluid in the metric. It is a generalisation parameter that affects the regularity of the black hole system. The Kretschmann scalar analysis, as done in Ref. \cite{Nascimento:2023tgw}, shows that the regularity of the Hayward solution changes for $\beta<-1$ and $\beta>0$, due to the presence of the fluid of strings. In other words, the Hayward-AdS black hole solution is regular in the interval $-1\leq\beta<0$.

Now, without losing generality, an isotropic static and spherically symmetric spacetime is expressed by the line element as
\begin{equation}
ds^2=e^\nu dt^2-e^\lambda dr^2-r^2 d\theta^2-r^2\sin^2\theta d\phi^2.
\label{eq:1.1}
\end{equation}

Therefore, in the presence of the cosmological constant and the source representing the fluid of strings, the Einstein field equations can be written as

\begin{equation}
e^{-\lambda}\left(\frac{\lambda'}{r}-\frac{1}{r^2}\right)+\frac{1}{r^2}= \frac{12l^2m^2}{(r^3+2l^2m)^{2}}+\Lambda-\frac{\varepsilon}{r^2}\left(\frac{b}{r}\right)^{2/\beta},\label{eq:1.17}
\end{equation}

\begin{equation}
-e^{-\lambda}\left(\frac{\nu'}{r}+\frac{1}{r^2}\right)+\frac{1}{r^2}= \frac{12 l^2 m^2}{(r^3+2l^2m)^{2}}+\Lambda-\frac{\varepsilon}{r^2}\left(\frac{b}{r}\right)^{2/\beta}, \label{eq:1.18}
\end{equation}

\begin{equation}
\frac{1}{2}e^{-\lambda}\left(\frac{\nu'\lambda'}{2}+\frac{\lambda'}{r}-\frac{\nu'}{r}-\frac{\nu'^2}{2}-\nu''\right)=\frac{3mq^2(2q^2-3r^2)}{(q^2+r^2)^{7/2}}+\Lambda+\frac{\varepsilon}{\beta r^2}\left(\frac{b}{r}\right)^{2/\beta}. \label{eq:1.19}
\end{equation}

\noindent Subtraction of Eq. (\ref{eq:1.17}) from Eq. (\ref{eq:1.18}) gives

\begin{equation}
\lambda=-\nu\Rightarrow\lambda'=-\nu'.
\label{eq:1.20}
\end{equation}

\noindent Addition of Eq. (\ref{eq:1.17}) with Eq. (\ref{eq:1.18}) using Eq. (\ref{eq:1.20}) simplifies as

\begin{equation}
e^{-\lambda}\frac{\lambda'}{r}-e^{-\lambda}\frac{1}{r^2}+\frac{1}{r^2}=\frac{12l^2m^2}{(r^3+2l^2m)^{2}}+\Lambda-\frac{\varepsilon}{r^2}\left(\frac{b}{r}\right)^{2/\beta}. \label{eq:1.21}
\end{equation}

Using the exact form of the metric functions $\nu$ and $\lambda$ as specified in Ref. \cite{Nascimento:2023tgw}, we can write

\begin{equation}
\nu=-\lambda=ln(1+f(r)).
\label{eq:1.22}
\end{equation}

Now, Eqs. (\ref{eq:1.21}) and (\ref{eq:1.19}) can be rewritten, considering Eqs. (\ref{eq:1.20}) and (\ref{eq:1.22}), respectively, as
 
\begin{equation}
-\frac{1}{r^2}(rf'+f)=\frac{12l^2m^2}{(r^3+2l^2m)^{2}}+\Lambda-\frac{\varepsilon}{r^2}\left(\frac{b}{r}\right)^{2/\beta},
\label{eq:1.25}
\end{equation}

\begin{equation}
2\frac{f'}{r}+f''=\frac{48(r^3-l^2m)l^2m^2}{(r^3+2l^2m)^{3}}-2\Lambda-\frac{2\varepsilon}{\beta r^2}\left(\frac{b}{r}\right)^{2/\beta}.
\label{eq:1.26}
\end{equation}
 Adding Eqs. (\ref{eq:1.25}) and (\ref{eq:1.26}) and multiplying the result by $r^2$, we get the following relation
 \begin{equation}
r^2f''+rf'-f+\Lambda r^2-\frac{12l^2m^2r^2}{(r^3+2l^2m)^{2}}-48\frac{(r^3-l^2m)l^2m^2r^2}{(r^3+2l62m)^{3}}+\varepsilon\left(\frac{\beta+2}{\beta}\right)\left(\frac{b}{r}\right)^{2/\beta}=0.\label{eq:1.27}
\end{equation}

The above differential equation can be solved to obtain the solution as

\begin{equation}
\begin{aligned}
f(r)=&-\frac{2 m r^2}{(r^3+2l^2m)}-\frac{\Lambda  r^2}{3}+\\
&\left\{\begin{array}{rcl}\varepsilon b[1+2\log (r)]/2r\;\;for\;\;\beta=2,\\
\varepsilon\beta(\beta -2)^{-1} \left(\frac{b}{r}\right)^{2/\beta}\;\;for\;\;\beta\neq 2.
\end{array}
\right.
\label{eq:1.28}
\end{aligned}
\end{equation}

\noindent Therefore, using Eq. (\ref{eq:1.28}) in Eq. (\ref{eq:1.22}) and then substituting it into Eq. (\ref{eq:1.1}), we can get metric function for the Hayward-AdS black hole surrounded by a fluid of strings as
\begin{equation}
ds^2=f(r) dt^2-f(r)^{-1} dr^2-r^2 d\theta^2-r^2\sin^2\theta d\phi^2.
\label{eq:line_element}
\end{equation}

It is expedient to redefine $ q^3\equiv 2l^{2}M$ to fix the product $ ml^{2} $ rather than the mass itself. This is a valid choice, avoiding any ambiguities. Without this redefinition, there would be two possible pressure values, one of which is not physically realistic. Therefore, incorporating these considerations, we finally arrive at the metric function for the black hole system cast as \cite{Nascimento:2023tgw, Waseem:2025bwb}
\begin{align}
f(r)&=1-\frac{2 M r^2}{r^3+q^3}-\frac{\Lambda  r^2}{3}+\varepsilon\beta(\beta -2)^{-1} \left(\frac{b}{r}\right)^{2/\beta}.
\label{eq:f_r}
\end{align}
The parameter $q$ here is analogous to the magnetic charge of the black hole and is related to the Hayward regularisation scale.

\section{Basic Thermodynamic Properties} \label{section3}
The inclusion of the cosmological parameter $\Lambda$ into the study of black hole thermodynamics makes it different from general thermodynamic considerations, as the black hole parameters are changed within a fixed AdS background. $\Lambda$ corresponds to the thermodynamic variable, pressure. Thus the black hole's volume is the conjugate variable of $\Lambda$. 
The relevant thermodynamic pressure \cite{Kubiznak:2012wp} can be expressed as 
\begin{equation}
    P = -\frac{\Lambda}{8\pi}. \label{pressure}
\end{equation}
The corresponding thermodynamic volume is given by
\begin{equation} \label{V0}
    V=\frac{4\pi r^3_+}{3} . 
\end{equation}

The first law of thermodynamics in the case of a charged, static black hole is obtained as
\begin{equation}
    dM = T_HdS + \Phi_H dq +V dP, \label{eq:first_law}
\end{equation}
where $M$ is the black hole's mass.

This gives the expressions for the thermodynamic temperature $T_H$, magnetic potential $\Phi_H$ and volume $V$, respectively, as
 \begin{equation}
 T_H = \frac{\partial M}{\partial S},\,\,\,\,\Phi_H = \frac{\partial M}{\partial q}\label{eq:TH_Ph},\,\,\,\,V = \frac{\partial M}{\partial P}.
\end{equation}
Using the condition $f(r_+) = 0$ in Eq. (\ref{eq:f_r}), the expression for $M$ can be obtained as 
\begin{equation}  
M=\frac{3 \beta\varepsilon  \left(\frac{b}{r_+}\right){}^{2/\beta }-\Lambda\beta r_+^2+3\beta+2 \Lambda  r_+^2-6}{6 (\beta -2) r_+^2\left(q^3+r_+^3\right)^{-1}}.
\label{eq:mass}
\end{equation}
Using the metric function in Eq. (\ref{eq:f_r}), and substituting the value of $M$ from Eq. (\ref{eq:mass}),  we can arrive at the expression for the Hawking temperature as
\begin{equation}
  T_H  = \frac{\kappa}{2\pi}= \frac{\beta \varepsilon  \left(\frac{b}{r_+}\right){}^{2/\beta } \left((\beta -2) r_+^3-2 (\beta +1) q^3\right)-(\beta -2) \beta  \left(2 q^3+\Lambda  r_+^5-r_+^3\right)}{4 \pi  (\beta -2) \beta  r_+ \left(q^3+r_+^3\right)},
  \label{Th}
\end{equation}
where,
\begin{equation}
    \kappa = \frac{f'(r)}{2}\Bigg{|}_{r=r_+}
\end{equation} is the surface gravity.

\begin{figure}[t!]
      	\centering{
      \includegraphics[scale=0.65]{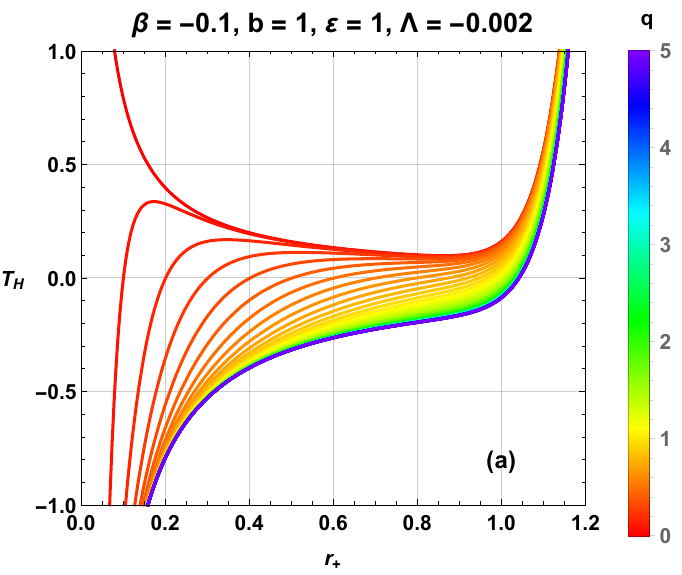} \hspace{10mm}
       \includegraphics[scale=0.65]{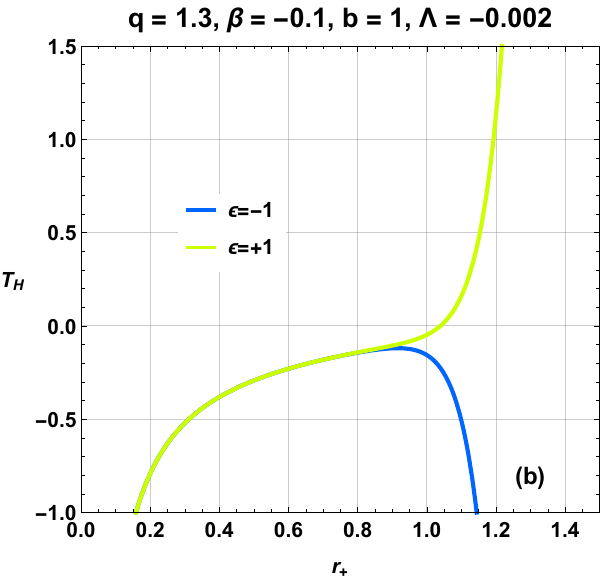}\\
     \vspace{0.4 cm}\hspace{.5cm}\includegraphics[scale=0.65]{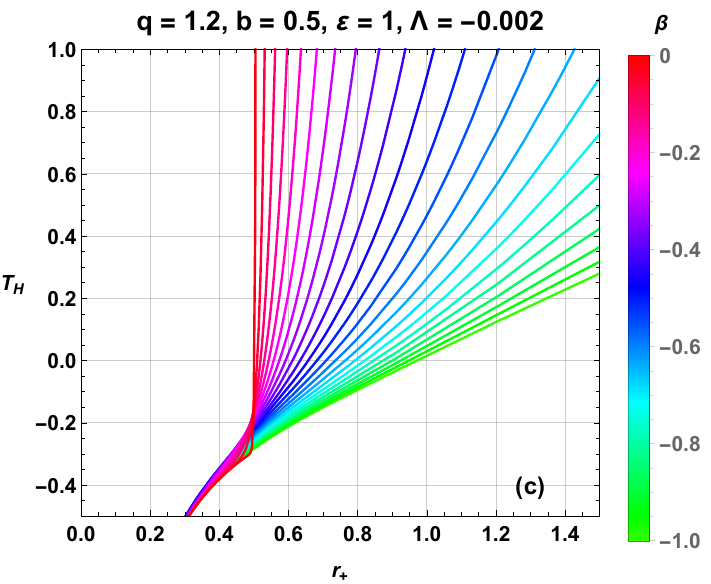}\hspace{7mm}
      \includegraphics[scale=0.65]{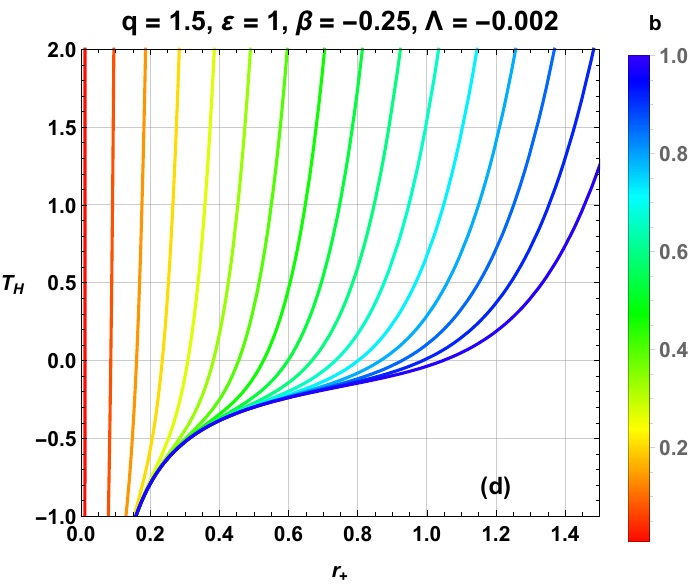}
       
       }
      	\caption{Variation of the Hawking temperature $T_H$ with the black hole horizon radius $r_+$.}
      	\label{figT01}
      \end{figure}
As clearly illustrated in Fig.\ref{figT01}, we depict the variation of the Hawking temperature $T_H$ with respect to the black hole horizon radius $r_+$ (using Eq. (\ref{Th})). The figure is divided into four panels, each highlighting the influence of different parameters on the temperature behaviour of the H-AdS black hole surrounded by a fluid of strings. Fig.\ref{figT01}(a) demonstrates the effect of the Hayward regulaisation parameter $q$ on $T_H$. Small values of $q$ correspond to negative temperatures as observed for small black holes (i.e., small $r_+$). This negative temperature indicates instability for small black holes under these conditions. As the horizon radius $r_+$ increases, the temperature undergoes a transition from negative to positive values, suggesting a stable thermodynamic phase for larger black holes. Raising $q$ causes the black hole metric to undergo a repulsive effect, changing its geometry and consequently the surface gravity at the horizon, which establishes $T_H$. The more $q$ grows, the more this parameter counteracts attractive gravity at the horizon.  As a result, $T_H$ for a given radius is lowered along with the surface gravity. Interestingly, for larger values of $q$, the temperature remains negative for comparatively larger black holes, indicating that the parameter $q$ significantly influences the stability and thermodynamic behaviour of the system, highlighting its critical role in determining the thermodynamic stability. In the case of $q = 0$, which represents a singular black hole, the temperature is always positive, regardless of the horizon radius. Fig.\ref{figT01}(b) portrays the influence of the energy density sign parameter $\varepsilon$, which is associated with the fluid of strings surrounding the black hole. It is observed that when $r_+$ is small, $T_H$ has negative values for both $\varepsilon=\pm 1$. This indicates that the black hole system is unstable when the size is small. As $r_+$ increases, the curves rise to a positive value of $T_H$. For $\varepsilon=+1$, the temperature keeps rising, which means that the positive string density makes the black hole hotter and thus more stable. The divergence seen in the curve may signal a possible phase transition. On the other hand, as we can see, the temperature becomes negative again near $r_+\approx1.1$, after initially rising for $\varepsilon=-1$. This means that the black hole with negative string density becomes unstable. Fig.\ref{figT01}(c) examines the effect of the parameter $\beta$ on $T_H$. Similar to the trends observed in the previous figures, smaller values of $\beta$ result in a negative temperature for smaller black holes. This further reinforces the idea that smaller black holes are more susceptible to thermodynamic instability when certain parameters, such as $\beta$, are small. As $r_+$ increase, the temperature becomes positive, indicating a transition to a stable phase. Fig.\ref{figT01}(d) shows the influence of $b$ on $T_H$. Similar to the other parameters, smaller values of $b$ lead to negative temperatures for smaller black holes. This indicates that $b$ also plays an important role in determining the thermodynamic stability of the system. As the horizon radius $r_+$ increases, the temperature undergoes transitions to positive values, suggesting that larger black holes are more stable regardless of the value of $b$.

A common trend across all four graphs in Fig. \ref{figT01} is that smaller values of the parameters $\varepsilon$, $\beta$, and $b$ lead to negative temperatures for smaller black holes. This suggests that these parameters play a crucial role in determining the thermodynamic stability of the system, particularly for black holes with smaller horizon radii. The smaller black holes are more sensitive to changes in these parameters, often leading to instability of the system, while larger black holes tend to exhibit stable thermodynamic behaviour. These findings provide valuable insights into the interplay between the black hole's properties and the surrounding fluid of strings, highlighting the conditions under which the system exhibits stable or unstable behaviours.

Again, from the first law of thermodynamics, the entropy can be obtained as
\begin{equation} \label{S0}
    S =\int \frac{1} {T_H} \, d 
    M= \frac{\pi  \left(r_+^3-2 q^3\right)}{r_+}.
\end{equation}
The enthalpy of the system is calculated using the standard formula
\begin{equation}\label{eq32}
     H = \int T_HdS.
\end{equation}
This gives the resulting enthalpy of the black hole system as
\begin{equation} \label{H0}
    H = \frac{1}{2} \left(\frac{\beta  \varepsilon  q^3 \left(\frac{b}{r_+}\right){}^{2/\beta }}{(\beta -2) r_+^2}+\frac{\varepsilon  r_+ \left(\frac{b}{r_+}\right){}^{2/\beta }}{1-\frac{2}{\beta }}-\frac{(2-\beta ) q^3}{(\beta -2) r_+^2}-\frac{\Lambda  r_+^3}{3}+r_+\right).
\end{equation}

The internal energy $(U)$ is a fundamental thermodynamic property that provides insight into the energy content of the system. For the H-AdS black hole surrounded by a fluid of strings, the internal energy is derived using the familiar relation $U=H-PV$. Using the expressions from Eqs. \eqref{V0}, \eqref{pressure} and \eqref{H0}, the internal energy reads as
\begin{equation}
     U = \frac{\left(q^3+r_+^3\right) \left(\beta  \varepsilon  \left(\frac{b}{r_+}\right){}^{2/\beta }+\beta -2\right)}{2 (\beta -2) r_+^2}.
\end{equation}
This expression highlights the dependence of the internal energy on the parameters $q$, $r_+$, $\beta$, $\varepsilon$, and $b$, which characterise the black hole and the surrounding fluid of strings. The parameters $\beta$, $\varepsilon$, and $b$ are associated with the fluid of strings, influencing the thermodynamic behaviour of the system.

In addition to the internal energy, the  Helmholtz free energy $(F)$ and specific heat $(C)$ are also crucial for understanding the stability and phase transitions of the system. The Helmholtz free energy provides information about the work obtainable from the system at constant temperature, while the specific heat indicates the system's response to changes in temperature. These quantities are derived using the thermodynamic relations and the previously obtained expressions for internal energy and other system properties.

In the subsequent sections, we investigate the effects of small stable fluctuations near equilibrium on the thermodynamic properties of the H-AdS black hole surrounded by a fluid of strings. These fluctuations are essential for understanding the stability of the system and its behaviour under perturbations. By analysing the variations in internal energy, Helmholtz free energy, and specific heat, we gain deeper insights into the thermodynamic stability and phase structure of the black hole system. This analysis contributes to the broader understanding of black hole thermodynamics and the role of external factors, such as a fluid of strings, in modifying the thermodynamic properties of black holes.

\section{Thermal Fluctuations and Corrected Thermodynamics}\label{section4}
In recent years, there has been growing interest in understanding the role of quantum corrections in black hole thermodynamics. These corrections arise due to thermal fluctuations and quantum effects that become significant as the black hole approaches its minimal possible size. While classical thermodynamics provides a robust framework for analyzing black hole entropy, it does not fully account for the quantum nature of spacetime. To address this, researchers have developed various approaches incorporating thermal fluctuations into the entropy formulation, leading to modified thermodynamic expressions that better capture the behaviour of small black holes. These corrections are particularly important in the context of AdS black holes, where holographic principles and quantum gravity considerations further motivate a deeper investigation into their statistical mechanics.

\subsection{Corrected entropy due to thermal fluctuations}
The works of Hawking and Page \cite{Hawking:1982dh} gave the revolutionary finding that a canonical ensemble framework can be used to model black holes in an asymptotically curved spacetime effectively. Applying this principle further, we consider the deformed AdS black hole as a system of canonical ensemble and analyze how the entropy of the black hole behaves under the influence of thermal fluctuations. We take the canonical ensemble to consist of $N$ particles, each of which has an energy spectrum $E_n$. The system's statistical behaviour is represented by the partition function, which is given as
\begin{equation}
    Z = \int_0^\infty  dE  \rho (E) e^{-\bar{\beta}_{\kappa} E},
\end{equation}
where $\bar{\beta}_{\kappa}$ corresponds to the inverse temperature (in units of the Boltzmann constant), and $\rho (E)$ denotes the canonical density of states corresponding to the average energy $E$. 
By using the partition function $Z$ and then applying Laplace inversion, we can determine the density of states as
\begin{equation}
    \rho (E) = \frac{1}{2 \pi i} \int^{\bar{\beta}_{0\kappa}+
i\infty}_{\bar{\beta}_{0\kappa} - i\infty} d \bar{\beta}_{\kappa}  e^{S(\bar{\beta}_{\kappa})},
\end{equation}
where the entropy is given by
\begin{equation}
    S = \bar{\beta}_{\kappa} E + \log Z.
\end{equation}
In the absence of all thermal fluctuations, $\bar{\beta}_{\kappa}$ gives the entropy at the equilibrium temperature, which is denoted as $\bar{\beta}_{0\kappa}$. However, a Taylor expansion around $\bar{\beta}_{0\kappa}$ results in the expression for the corrected entropy in the presence of thermal fluctuations given as
\begin{equation}
    S = S_0 + \frac{1}{2}(\bar{\beta}_{\kappa} - \bar{\beta}_{0\kappa})^2 \left(\frac{\partial^2 S(\bar{\beta}_{\kappa})}{\partial \bar{\beta}_{\kappa}^2 }\right)_{\bar{\beta}_{\kappa} = \bar{\beta}_{0\kappa}} + \frac{1}{6}(\bar{\beta}_{\kappa} - \bar{\beta}_{0\kappa})^3 \left(\frac{\partial^3 S(\bar{\beta}_{\kappa})}{\partial \bar{\beta}_{\kappa}^3 }\right)_{\bar{\beta}_{\kappa} = \bar{\beta}_{0\kappa}} + \cdots,\label{39}
\end{equation}
here, the dots denote higher-order corrections.

It is important to note that, at equilibrium temperature, the first derivative of the entropy for $\bar{\beta}_{\kappa}$ becomes zero, which consequently leads to the expression for the density of states as
\begin{equation}
    \rho (E) = \frac{e^{S_0}}{2 \pi i} \int^{\bar{\beta}_{0\kappa} + i\infty}_{\bar{\beta}_{0\kappa} - i\infty} d \bar{\beta}_{\kappa} \, \exp \left[ \frac{(\bar{\beta}_{\kappa} - \bar{\beta}_{0\kappa})^2}{2} \left(\frac{\partial^2 S(\bar{\beta}_{\kappa})}{\partial \bar{\beta}_{\kappa}^2 }\right)_{\bar{\beta}_{\kappa} = \bar{\beta}_{0\kappa}} + \frac{(\bar{\beta}_{\kappa} - \bar{\beta}_{0\kappa})^3}{6} \left(\frac{\partial^3 S(\bar{\beta}_{\kappa})}{\partial \bar{\beta}_{\kappa}^3 }\right)_{\bar{\beta}_{\kappa} = \bar{\beta}_{0\kappa}} + \cdots \right].
\end{equation}

This approach, as mentioned in Refs. \cite{Gogoi:2024ypn,More:2004hv}, is used to simplify the entropy expression from Eq. (\ref{39})
\begin{equation}
    S = S_0 - \frac{1}{2} \log{S_{0}T_{h}^{2}} + \frac{f(m,n)}{S_{0}} + \cdots,
\end{equation}
where $f(m,n)$ is a constant. The expression for corrected entropy is further generalised with the help of Ref. \cite{Pourhassan:2016zzc} as
\begin{equation}
    S = S_0 - \lambda_1 \log(S_0 T_{H}^2) + \frac{\lambda_2}{S_0} + \cdots.
\end{equation}
The parameters $\lambda_1$ (quantum correction parameter) and $\lambda_2$ (geometric correction parameter) are defined to account for the corrected entropy terms of the first-order and second-order, respectively. The quantum/thermal fluctuations are the cause of the first-order term, which is the principal correction to the classical area law.  Usually, it is logarithmic. The fact that the horizon is not a perfectly rigid surface is captured by the first-order correction term.  The count of states is somewhat altered by tiny oscillations or quantum fields close to the horizon, and the effect is proportional to the logarithm of the black hole's size. The second-order entropy correction follows the logarithmic correction.  It quantifies how the number of black hole microstates is altered by higher-loop effects.  It only matters when the black hole is small or extremely near to an external state because it falls off as an inverse power of the area (or entropy). For $\lambda_1 \rightarrow 0$ and $\lambda_2 \rightarrow 0$, the original results can be recovered. Also, $\lambda_1 = 1$ and $\lambda_2 = 0$ produce the usual corrections  \cite{More:2004hv, Sadeghi:2016dvc}. The thermal fluctuations play a crucial role in altering the thermodynamics of black holes, becoming increasingly significant as their size diminishes.
\begin{figure}[h!]
      	\centering{
      \hspace{-.7cm}\includegraphics[scale=0.65]{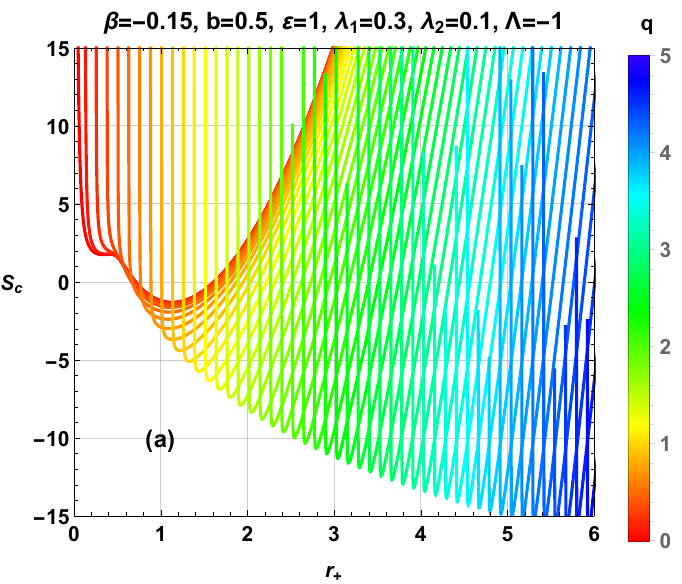}\hspace{5mm}
       \includegraphics[scale=0.65]{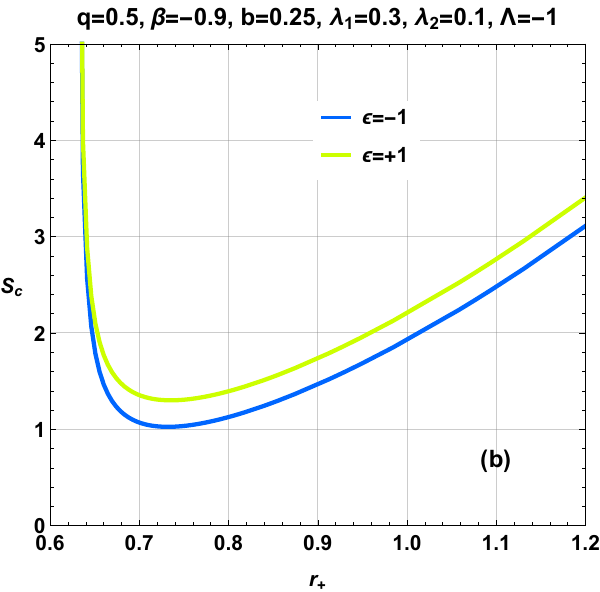}\\
      \vspace{0.4 cm}\includegraphics[scale=0.65]{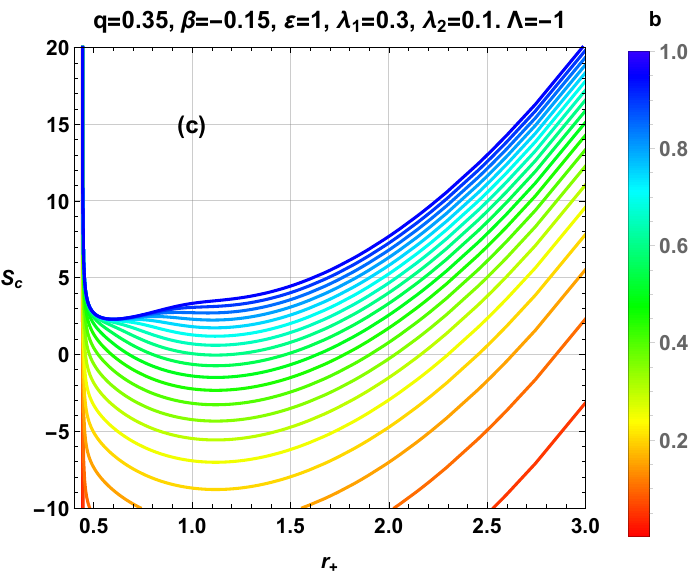}\hspace{5mm}
      \includegraphics[scale=0.65]{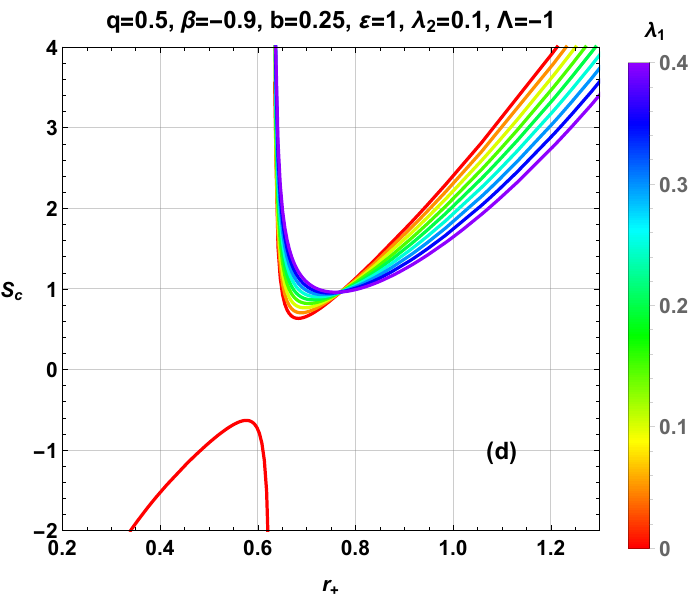}\\
      \vspace{0.4 cm}\includegraphics[scale=0.62]{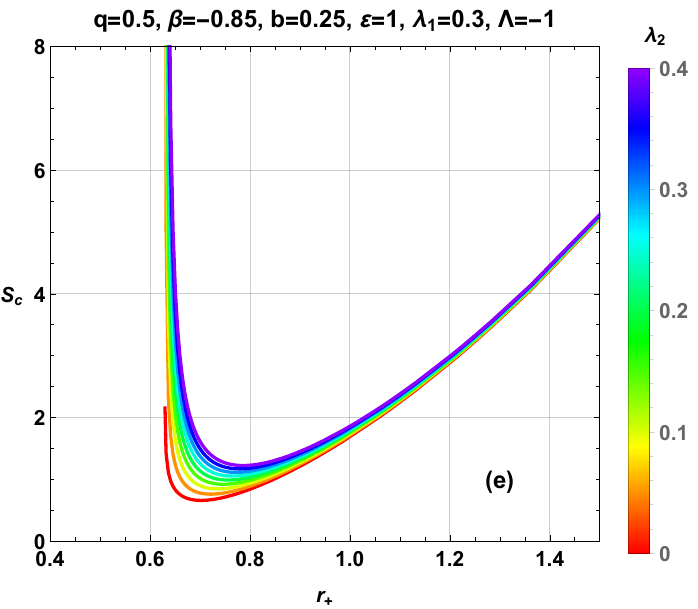}\hspace{5mm}
      \includegraphics[scale=0.65]{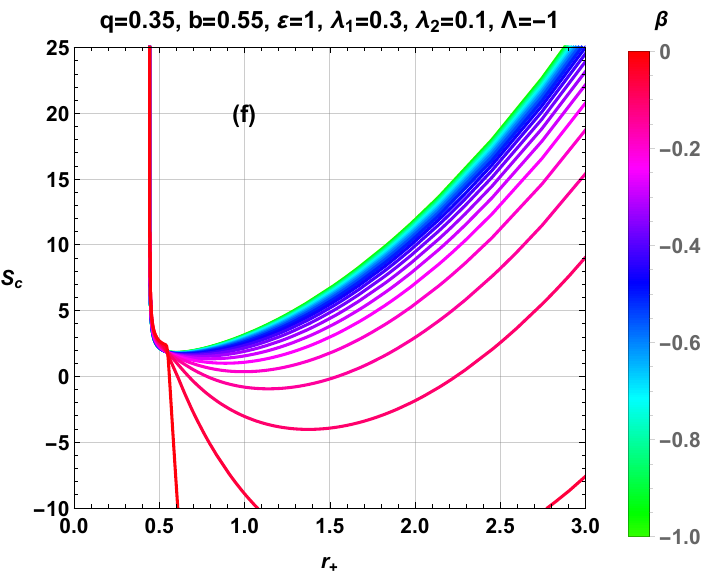}
       }
      	\caption{Variation of the corrected black hole entropy ($S_c$) with the horizon radius ($r_+$)}
      	\label{figSc01}
      \end{figure}

Including the second-order correction terms, the resultant entropy of the  black hole system can be explicitly given as  
\begin{multline}
    S_c = \pi  \left(r_+^2-\frac{2 q^3}{r_+}\right)+\lambda _1 \log (16 \pi )+\frac{\lambda _2 r_+}{\pi  \left(r_+^3-2 q^3\right)} \\ -\lambda _1 \log \left[\frac{\left(r_+^3-2 q^3\right) \left(\varepsilon  \left(\frac{b}{r_+}\right){}^{2/\beta } \left(2 (\beta +1) q^3-(\beta -2) r_+^3\right)+(\beta -2) \left(2 q^3+\Lambda  r_+^5-r_+^3\right)\right){}^2}{(\beta -2)^2 r_+^3 \left(q^3+r_+^3\right){}^2}\right].
    \label{sc}
\end{multline}
In Fig.\ref{figSc01}, we represent the black hole entropy as a function of its horizon radius with the help of Eq.(\ref{sc}), incorporating quantum corrections, charge, and cosmological effects. In Fig.\ref{figSc01}(a), the entropy $S_c$ decreases to a minimum and then increases as $r_+$ grows. The oscillatory behaviour of the entropy at higher values of $r_+$ accounts for some underlying fluctuations. These fluctuations may arise due to quantum corrections or instability regions, implying the existence of an unstable mode. The value of $q$ significantly affects the corrected entropy $S_c$. In the red/yellow region where $q$ is small, the decrease in entropy is steeper at small $r_+$. On the other hand, for larger values of $q$ (green to blue regions), the oscillations of entropy become more prominent. In the scenario of quantum gravity, the oscillations in $S_c$ may indicate the phase transitions or black hole stability conditions. The minimum entropy point can represent a phase transition in the black hole's evolution or a thermodynamic threshold. In Fig.\ref{figSc01}(b), the behaviour of corrected entropy $S_c$ as a function of $r_+$ for different values of $\varepsilon$ is depicted.  For both $\varepsilon=\pm1$, $S_c$ shows a U-shaped behaviour. It starts at a higher value for small $r_+$, reaches a minimum around $r_+\approx0.7$, and then starts increasing with $r_+$. $S_c$ is consistently higher when the string energy density is positive. This means that when $\varepsilon$ is positive, the number of microscopic states is higher and thus the thermodynamic stability is enhanced. Fig.\ref{figSc01}(c) shows the dependence of $S_c$ on $b$. Here also, $S_c$ shows similar behaviuor - it initially decreases to a minimum and then increases as $r_+$ increases. The red regions show that $S_c$ is lower for small $b$. Moving towards the blue region, $S_c$ increases rapidly with increasing $b$. There are no spikes or discontinuities in the curves, indicating a stable evolution of entropy. In Fig.\ref{figSc01}(d), $S_c$ shows a more complex behaviour. It has discontinuities and a sharp dip at small $r_+$, possibly indicating phase transitions or instability regions. Red regions, where $\lambda_1$ is small, have relatively smaller $S_c$. As $r_+$ increases (blue regions), the function increases with more stability. Also, a small peak is noticed at very low $r_+$. This might indicate a distinct thermodynamic phase. Fig.\ref{figSc01}(e) shows that $S_c$ is initially very high for small $r_+$ and then decreases sharply as $r_+$ increases for different values of $\lambda_2$. This initial decrease in entropy again hints at a phase transition or quantum corrections.  Beyond the point $r_+ = 0.6$, $S_c$ stabilises and shows a gradual increase. This shows that the normal thermodynamic behaviour starts to dominate for larger black holes. The closely spaced curves for different $\lambda_2$ values indicate that this parameter has almost negligible impact on $S_c$. From Fig.\ref{figSc01}(f) we can see that when $\beta$ is close to $0$, $S_c$ becomes negative. Negative entropy suggests an unstable state. The black hole area law does not allow the presence of negative entropy. This unusual behaviour is possibly due to the inclusion of the correction parameters. $S_c$ is positive beyond $r_+=0.5$ for smaller values of $\beta$, after which the corrected entropy increases sharply, a behaviour similar to those we have already observed for the other parameters. This implies that the system is getting closer to a more traditional thermodynamic regime for larger black holes. This also shows a significant impact of the $\beta$ on $S_c$. The red curves suggest that at $\beta = 0$,  $S_c$ is negative. With more negative values of $\beta$, $S_c$ starts to increase. Larger black holes with more substantial entropy are possible with this set of parameters. 

\subsection{Corrected Potentials Due to Thermal Fluctuations}
In this section, we will deal with various corrected thermodynamic variables for the H-AdS black hole surrounded by a fluid of strings. Considering the corrected entropy we have discussed in the previous section, we find the corrected enthalpy $H_c$ as
\begin{equation}
    H_c= \int{T_H dS_c} + \int{V_c dP}  \label{Enthalpy}.
\end{equation}
Here, $V_c$ denotes the corrected volume of the black hole system. We can neglect the second term of Eq. (\ref{Enthalpy}), because this term has no contribution to a constant value of $\Lambda$. Using $T_H$ and $S_c$ from Eqs. (\ref{eq:TH_Ph}) and (\ref{sc}) respectively, we obtain the expression for the corrected enthalpy, which is given in the appendix of this paper (Eq.(\ref{Hc})). 
\begin{figure}[t!]
      	\centering{
      \includegraphics[scale=0.65]{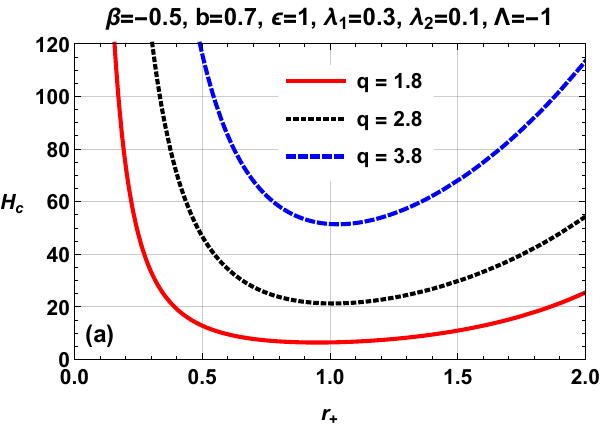}\hspace{5mm}
      \includegraphics[scale=0.65]{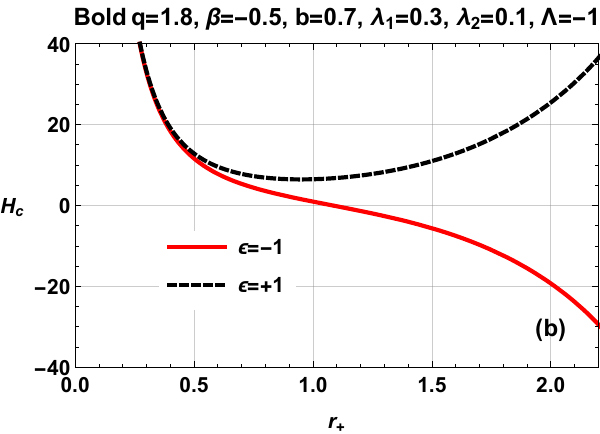}\\
      \vspace{0.4 cm}\includegraphics[scale=0.65]{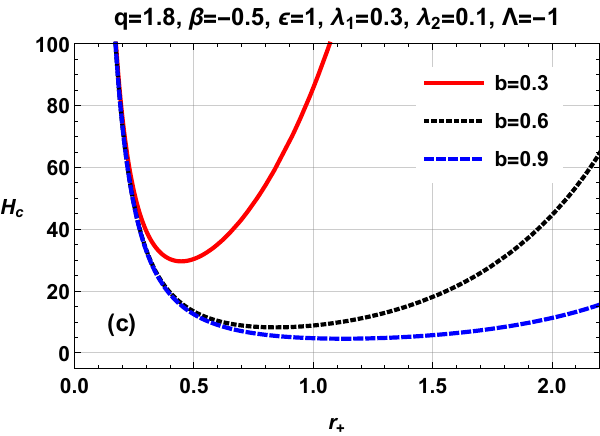}\hspace{5mm}
      \includegraphics[scale=0.65]{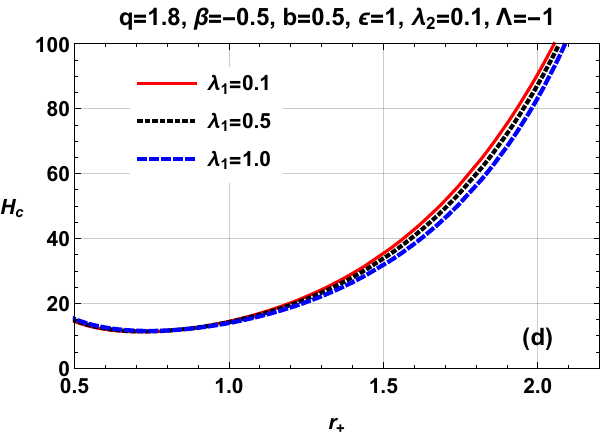}\\
      \vspace{0.4 cm}\includegraphics[scale=0.65]{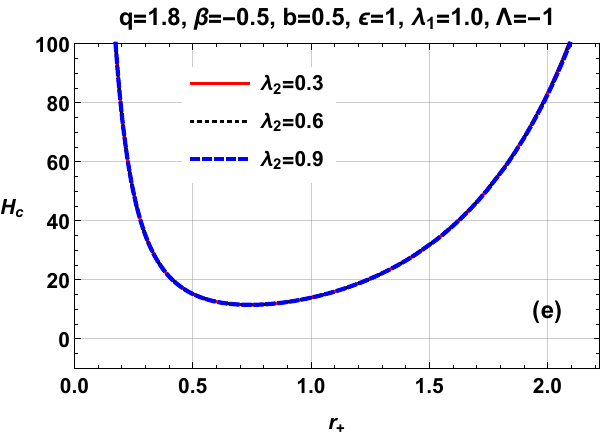}\hspace{5mm}
      \includegraphics[scale=0.65]{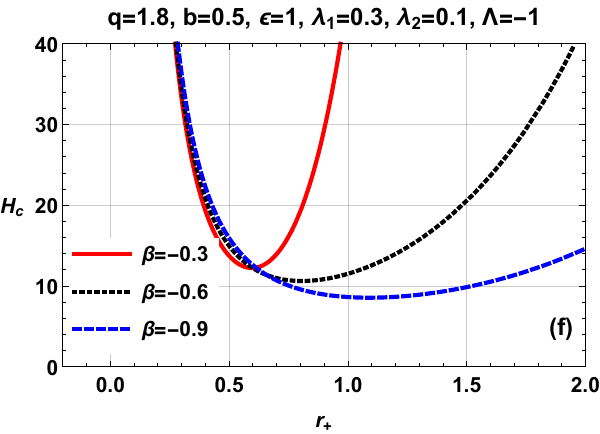}
       }
      	\caption{Variation of the corrected black hole enthalpy $H_c$ with the horizon radius $r_+$.}
      	\label{hc01}
      \end{figure}    
As in Fig.\ref{hc01}, we show a graphical representation of the dependence of the corrected enthalpy $H_c$ on the horizon radius $r_+$ with varying values of the model parameters (using Eq(\ref{Hc})). In Fig.\ref{hc01}(a), we can see that, when $q$ is higher, at the same $r_+$ the system has higher $H_c$. This indicates a positive contribution of $q$ to $H_c$. There is a minimum or inflection point on the graph, signifying a critical radius $r_c$ at which the black hole changes from a stable state to an unstable state. In AdS black holes having pressure, this is typical. In the larger radius regions, $H_c$ grows rapidly. This happens due to the pressure term and the modifications introduced by the correction parameters. Also, the positive slope of the curves in this region suggests that the black hole system is thermodynamically stable. However, the negative slope of the curves in the smaller radius region suggests instability, which may lead to phase transitions. In Fig.\ref{hc01}(b), the dependence of $H_c$ on $r_+$ for two different values of $\varepsilon$ is shown. For $\varepsilon=+1$, $H_c$ first decreases at small $r_+$ to a minimum and then starts increasing. Positive enthalpy values indicate that the black hole system is stabilised for positive $\varepsilon$ values. On the other hand, when this parameter obtains a negative value, i.e., $\varepsilon=-1$, enthalpy decreases with an increase of $r_+$ and becomes negative at larger $r_+$ values. This is a clear indication that negative $\varepsilon$ values make the sytem thermodynamically unstable beyond a certain size. Therfore, $\varepsilon$ may have greater control over the thermodynamic behaviour of our system. From Fig.\ref{hc01}(c), we see that the variation of $H_c$ due to $b$ is similar to that due to $q$, where the $H_c$-curve shifts more upward when $b$ is smaller, meaning that its increase leads to black holes having lower energy content. In Fig.\ref{hc01}(d), we see that increasing the quantum correction parameter $\lambda_1$ slightly lowers the curve, making black holes somewhat less energetic. In Fig.\ref{hc01}(e), it is observed that the secondary correction parameter $\lambda_2$ has no dramatic effect on $H_c$ at this scale, as the curves almost overlap. Slight deviations are noticed only at large $r_+$. The Fig.\ref{hc01}(f), shows the variation of $H_c$ due to the parameter $\beta$. As $\beta$ becomes more negative, the minimum enthalpy increases and the curves shift towards smaller $r_+$. $\beta$ has a significant impact on larger black holes. 
An expression for $V_c$ can be derived using $H_c$ from Eq.(\ref{Hc}) as 
\begin{align}
    V_c=\frac{2}{9} \left(\frac{\sqrt[3]{2} \left(2 \sqrt{3} \left(3 \pi  \lambda _1 q^2-\sqrt[3]{2} \lambda _2\right) \tan ^{-1}\left(\frac{q+2^{2/3} r_+}{\sqrt{3} q}\right)-\left(\sqrt[3]{2} \lambda _2+3 \pi  \lambda _1 q^2\right) V_{\text{t2}}\right)}{\pi  q}+6 r_+ V_{\text{t1}}\right),
    \label{V_c}
\end{align}
where, $V_{\text{t1}}=\frac{\lambda _2 r_+}{\pi  \left(r_+^3-2 q^3\right)}+3 \lambda _1 \left(\frac{q^3}{q^3+r_+^3}-2\right)+\pi  r_+^2$ and $V_{\text{t2}}=2 \log \left(2 q-2^{2/3} r_+\right)-\log \left(2 q^2+\sqrt[3]{2} r_+ \left(\sqrt[3]{2} q+r_+\right)\right).$
As we can see, due to the presence of quantum fluctuations, the volume of the black hole would also be affected. 

Now, let us find out how the thermal fluctuations affect the Helmholtz free energy $F$. The uncorrected expression for $F$ is given by
\begin{equation}
    F = \frac{1}{4} \left[\frac{\varepsilon  \left(\frac{b}{r_+}\right){}^{2/\beta } \left(-2 (\beta +2) q^6+2 (4 \beta -1) q^3 r_+^3+(\beta +2) r_+^6\right)}{(\beta -2) r_+^2 \left(q^3+r_+^3\right)}-\frac{2 q^3}{r_+^2}+\frac{3 \Lambda  q^6+9 q^3 r_+}{q^3+r_+^3}+\Lambda  r_+^3+r_+\right].
    \label{F}
\end{equation}
When thermal fluctuations are introduced, this expression is corrected using the standard relation
\begin{equation}
    F_c = -\int{S_c dT_H}- \int{PdV_c} .
\label{Fc}
\end{equation}
\begin{figure}[t!]
      	\centering{
        \includegraphics[scale=0.65]{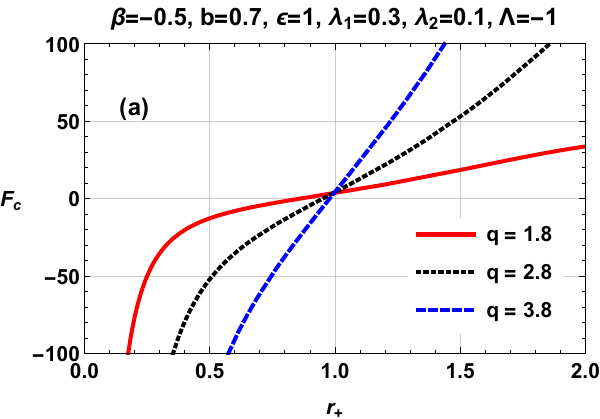} \hspace{5mm}
        \includegraphics[scale=0.65]{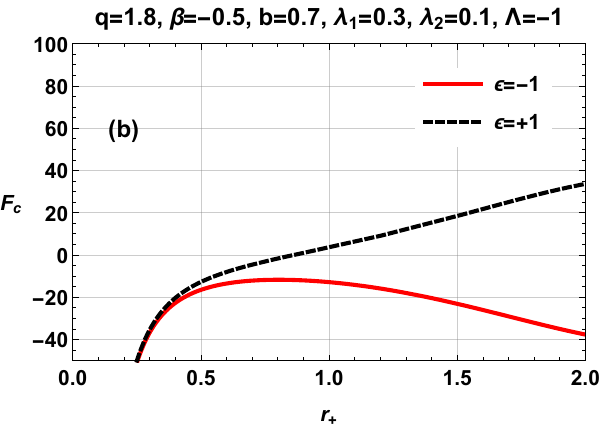}\\
        \vspace{0.4 cm}\includegraphics[scale=0.65]{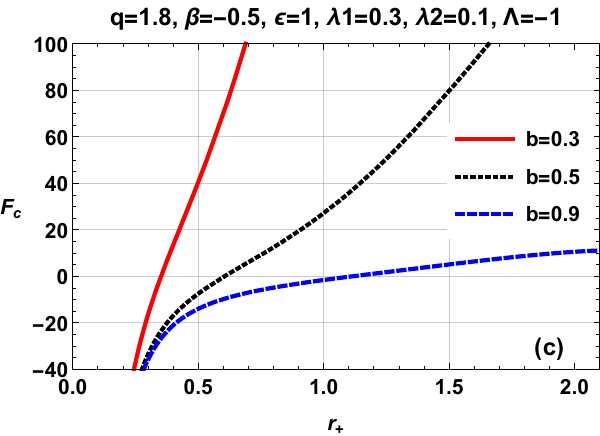}\hspace{5mm}
        \includegraphics[scale=0.65]{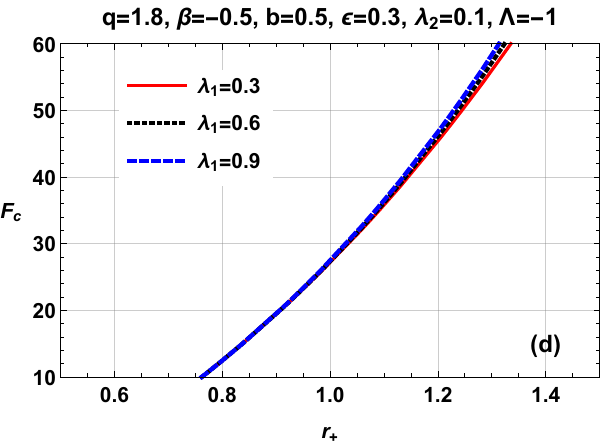}\\
        \vspace{0.4 cm}\includegraphics[scale=0.65]{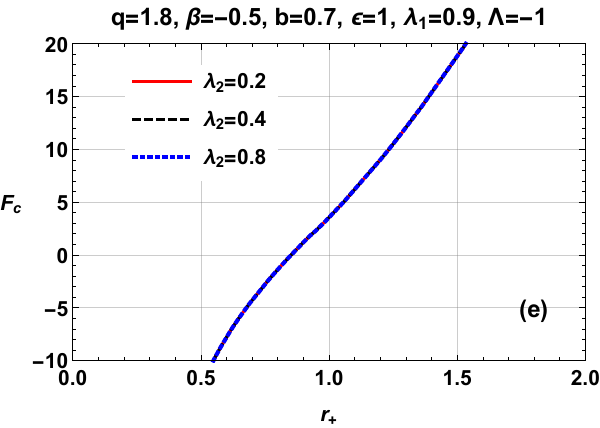}\hspace{5mm}
        \includegraphics[scale=0.65]{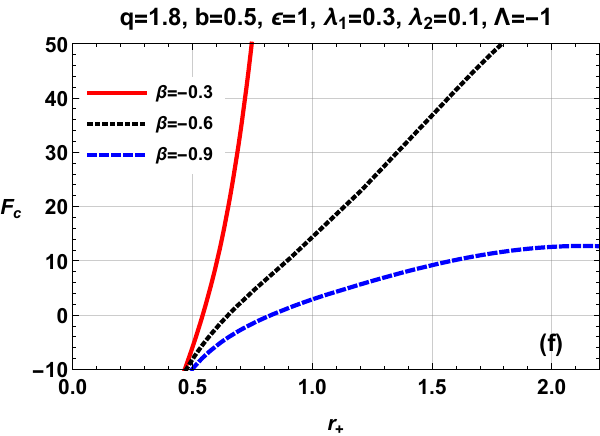}
       }
      	\caption{Variation of corrected black hole Helmholtz free energy ($F_c$) with the horizon radius ($r_+$).}
      	\label{fc}
      \end{figure}
For the explicit expression of $F_c$, readers may see the Section \ref{appen} of the paper. The Eq.(\ref{Fc_explicit}) is used to physically interpret the variation of corrected Helmholtz free energy as shown in Fig.\ref{fc}. We can see from Fig.\ref{fc}(a) that, at small $r_+$, where quantum corrections are significant, $F_c$ drops sharply. This indicates instability or a thermodynamically unfavourable system in that regime. Small black holes are unstable compared to larger ones, however, the stability of small black holes increases with high $q$. This is due to the fact that, the regularisation parameter $q$ provides repulsive energy, which can offset the tendency for collapse at small radii. Larger $q$ makes $F_c$ decrease faster. The point $F_c=0$ indicates a phase transition. The variation of $F_c$ for $\varepsilon$ in Fig.\ref{fc}(b) shows that at the small $r_+$ region, $F_c$ has negative values for both $\varepsilon=\pm1$, which implies that small black holes are thermodynamically favourable. For $\varepsilon=+1$, $F_c$ increases with $r_+$. It becomes positive for large $r_+$, which means thermodynamic instability at large horizon radius. Again, for $\varepsilon=-1$, $F_c$ first increases and then starts to decrease after reaching a maximum. This behaviour points to the possibility of a phase transition as $r_+$ varies. Fig.\ref{fc}(c) shows the dependence of $F_c$ on $b$. It is observed that as $b$ increases, $F_c$ becomes more negative and black holes stabilise at larger radii. It suggests that a higher $b$ favours larger and more stable black holes with lower free energy. A similar behaviour is also observed in the case of the variation of $F_c$ due to $\lambda_1$ (Fig.\ref{fc}(d)), where increasing $\lambda_1$ raises the $F_c$ curves. It makes the black hole less favorable thermodynamically for larger $\lambda_1$. However, the overlapping curves show that this parameter does not have a significant impact on $F_c$ at lower radii. Fig.\ref{fc}(e) shows that increasing $\lambda_2$ also raises $F_c$ slightly, but changes are small. It means that $\lambda_2$ has a minor effect, slightly reducing thermodynamic stability at higher values. Fig.\ref{fc}(f) shows that as $\beta$ becomes more negative, $F_c$ decreases and curves drop lower. This behaviour suggests that more negative $\beta$ values enhance black hole stability by lowering the free energy.

The corrections introduced by quantum fluctuations affect other thermodynamic parameters, such as specific heat, internal energy, Gibbs free energy, etc. The corrected internal energy expression is derived using the formula
\begin{equation}
    U_c=H_c-P V_c.
    \label{uc}
\end{equation}
Using the previously derived values of $H_c$, $P$ and $V_c$, we can obtain the explicit form of corrected internal energy $U_c$. Readers are requested to see the appendix of this paper for the expression of corrected internal energy (Eq.\ref{Uc}).

\begin{figure}[t!]
      	\centering{
        \includegraphics[scale=0.65]{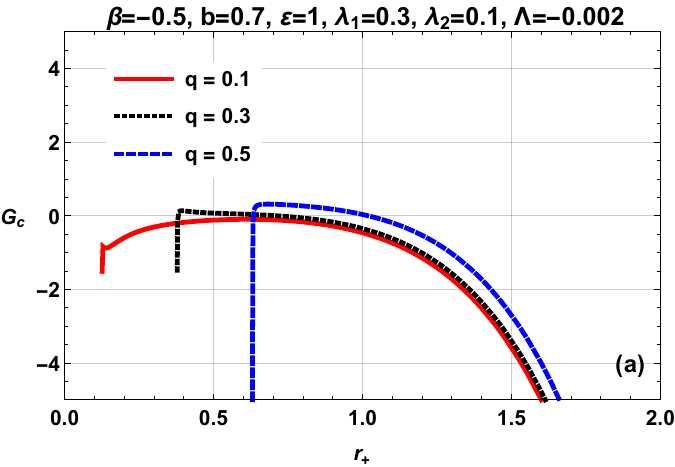}\hspace{5mm}
       \includegraphics[scale=0.65]{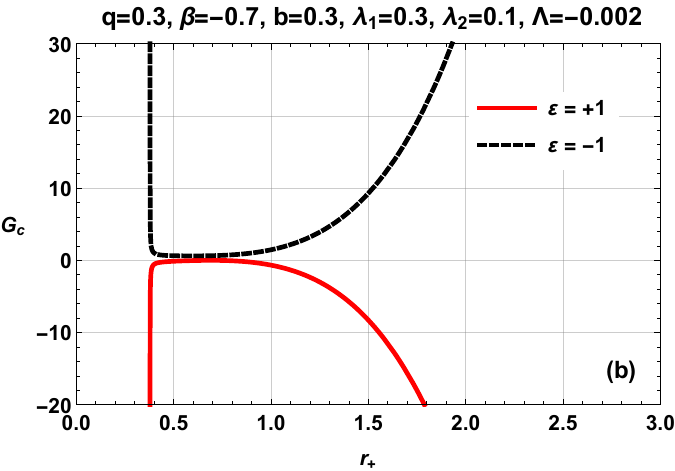}\\
       \vspace{0.4 cm}\includegraphics[scale=0.65]{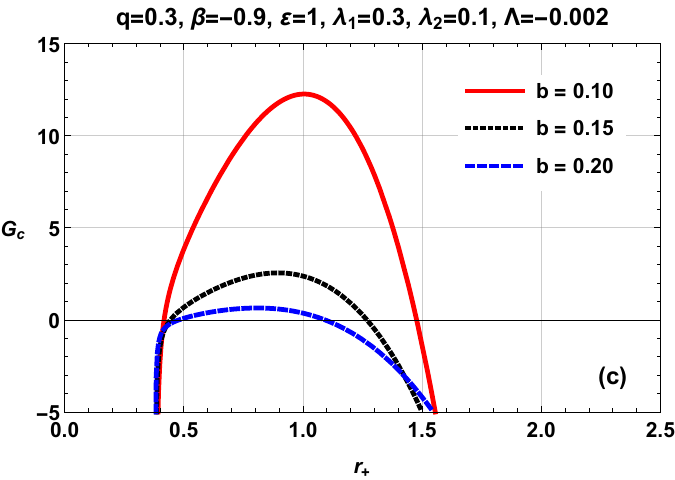}\hspace{5mm}
       \includegraphics[scale=0.65]{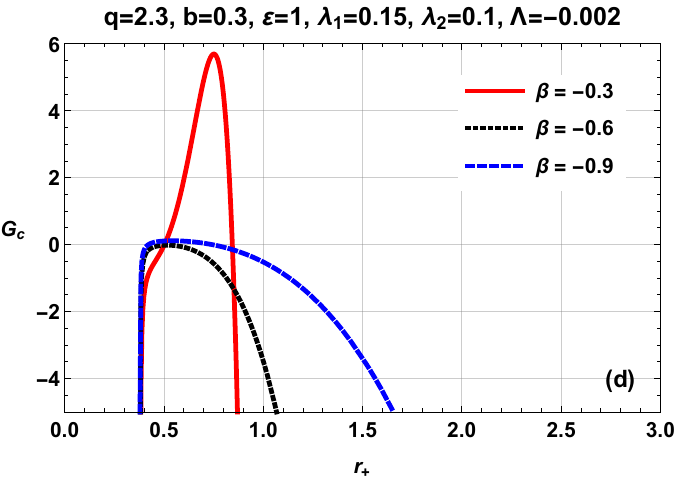}\\
       \vspace{0.4 cm}\includegraphics[scale=0.65]{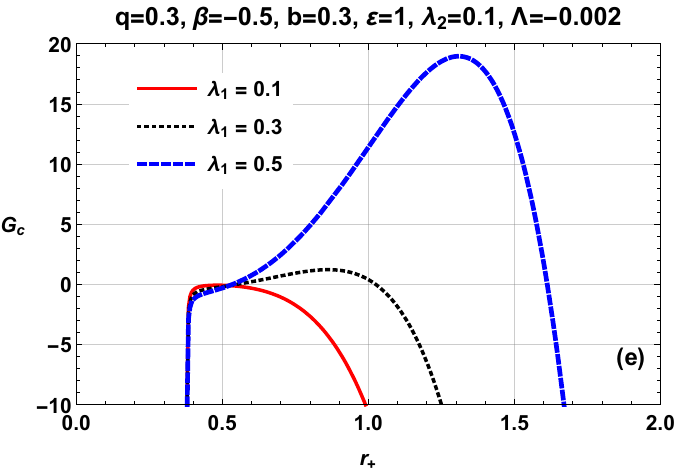}\hspace{5mm}
       \includegraphics[scale=0.65]{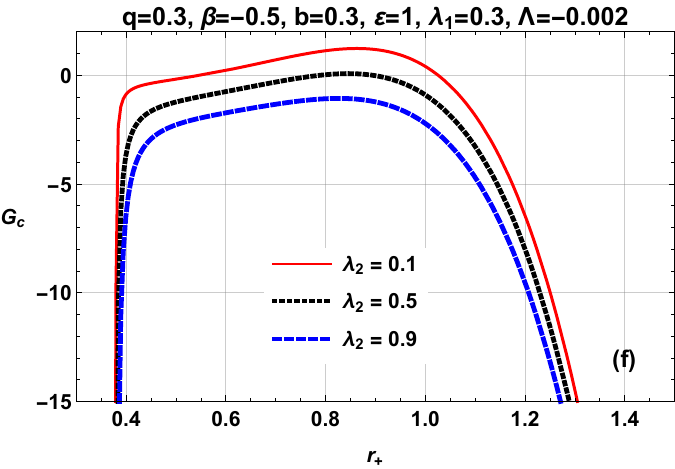}
       }
      	\caption{Variation of the black hole $G_c$ with the horizon radius $r_+$.}
      	\label{figGc01}
      \end{figure}

The equation for determining the corrected Gibbs free energy is
\begin{equation}
    G_c=M - T_H S_c.
\end{equation}
Using the corrected entropy from Eq. (\ref{sc}) we get the expression for $G_c$ as
\begin{align}
G_c =\; & -\frac{1}{12 (\beta - 2) r_+^2 (q^3 + r_+^3)} \bigg[ 
    \frac{3 r_+}{\pi} 
    \left( 
        \phi_0 - \phi_1
    \right)
    \left( 
        -\lambda_1 \log \left( 
            \frac{(r_+^3 - 2 q^3) 
            \left( 
                -\phi_0 
                + \phi_1 
            \right)^2} 
            {(\beta - 2)^2 r_+^3 (q^3 + r_+^3)^2} 
        \right) 
        + \lambda_1 \log (16 \pi)
          \right. \nonumber \\
& \left.
        + \frac{\lambda_2 r_+}{\pi (r_+^3 - 2 q^3)} 
        + \pi \left( r_+^2 - \frac{2 q^3}{r_+} \right) 
    \right) + 2 (q^3 + r_+^3)^2 
    \left( 
        (\beta-2)(\Lambda r_+^2 - 3) 
        - 3 \beta \varepsilon \left( \frac{b}{r_+} \right)^{2/\beta} 
    \right) 
\bigg],\label{Gc}
\end{align}
where 
$\phi_0 = \varepsilon \left( \frac{b}{r_+} \right)^{2/\beta} 
        \left( (\beta - 2) r_+^3 - 2 (\beta + 1) q^3 \right)$
and 
$\phi_1= (\beta - 2) (2 q^3 + \Lambda r_+^5 - r_+^3)$.
Hence, we can see that the thermal fluctuations change the expression for the Gibbs free energy. This means that they affect the maximum amount of mechanical work obtainable from the system. A plot of $G_c$ showing its variations with respect to the first and second order correction parameters, which is obtained in Eq.(\ref{Gc}), is shown in Fig.\ref{figGc01}. 
Figs.\ref{figGc01}(a) shows the variation of $G_c$ for the parameter $q$. From here, we have observed that $G_c$ grows in general for the same $r_+$ as $q$ increases. $G_c$  is seen to have negative values, which often signals a thermodynamically favoured phase.  A higher $q$ causes the graph to go upward, suggesting that a higher charge tends to make the system more stable. The dependence of $G_c$ on $r_+$ for two different values of $\varepsilon$ is shown in Fig.\ref{figGc01}(b). For $\varepsilon=+1$, $G_c$ decreases with increasing $r_+$, obtaining more negative values, and then hits a minimum. It may indicate a thermodynamically preferred state. For $\varepsilon=-1$, $G_c$ has large positive values in the small $r_+$ regions, reaches a minimum and finally starts to increase for large $r_+$ values.  Fig.\ref{figGc01}(c) and Fig.\ref{figGc01}(d) show the variation of $G_c$ for parameters $b$ and $\beta$, respectively. For small $b$, $G_c$ shows an extreme behaviour. The pronounced maximum and minimum are typical of a first-order phase transition. With an increase in $b$, this behaviour smooths out, suggesting that the parameter $b$ has some control over the strength of phase transitions. Similarly, $G_c$ has a peak value corresponding to each  $\beta$ value. As $\beta$ becomes more negative, this peak shifts to the right. Also, the magnitude of the peak decreases, making the curves more flattened. This signifies that a more negative value of $\beta$ makes the system thermodynamically more stable. Fig.\ref{figGc01}(e) and Fig.\ref{figGc01}(f) show the behaviour of $G_c$ for different values of the correction parameters $\lambda_1$ and $\lambda_2$. From here, we observe that, when the value of $\lambda_1$ increases, the peak in the $G_c$ curve becomes more positive. This may be an indication that smaller $\lambda_1$ has greater contribution to the stability of the system. The critical point behaviour becomes clearer for large $\lambda_1$ values. This may point towards a possible phase transition of the black hole system. Similar to this behaviour, as $\lambda_2$ increases, the $G_c$ curves flatten. There is a slight shift in the peak value. This indicates that higher $\lambda_2$ values may contribute to greater stability of the black hole. 

\section{Phase Transition and Stability}
\label{section5}
\begin{figure}[t!]
      	\centering{
       \includegraphics[scale=0.65]{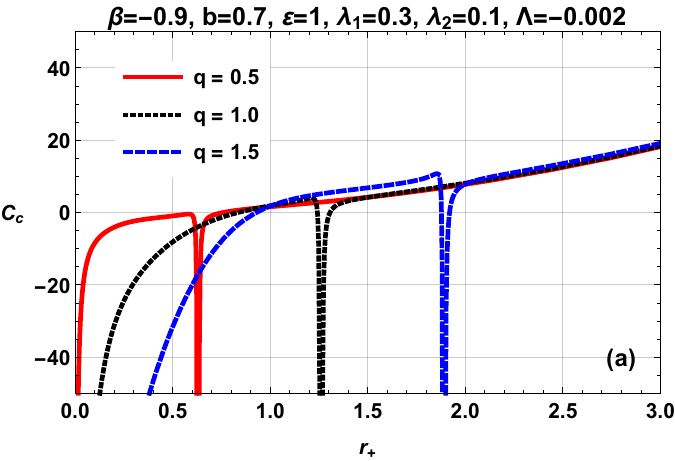}\hspace{5mm}
       \includegraphics[scale=0.65]{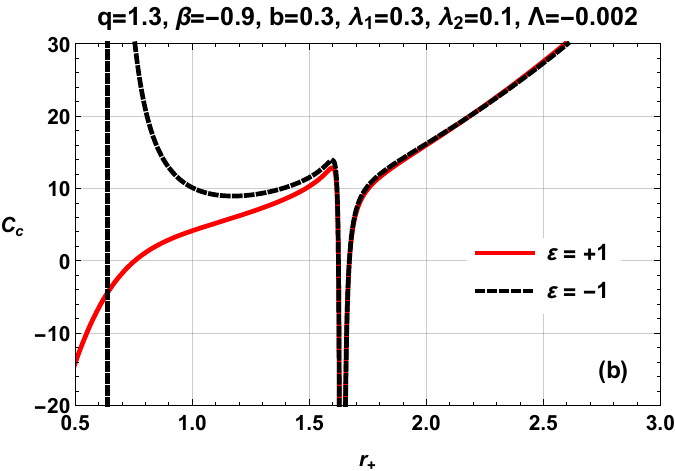}\\
       \vspace{0.4 cm}\includegraphics[scale=0.65]{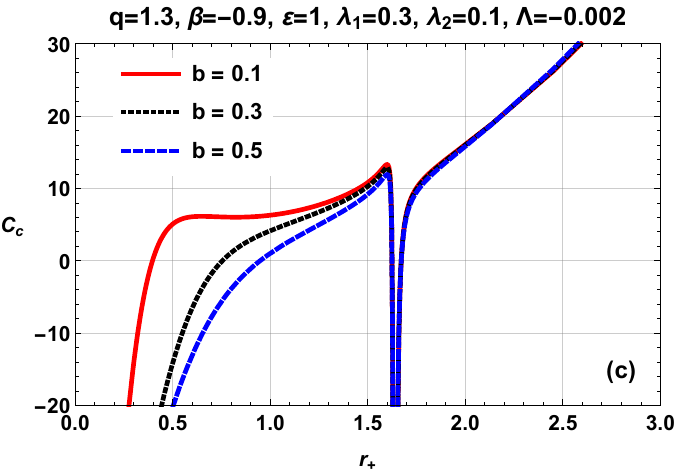}\hspace{5mm}
       \includegraphics[scale=0.65]{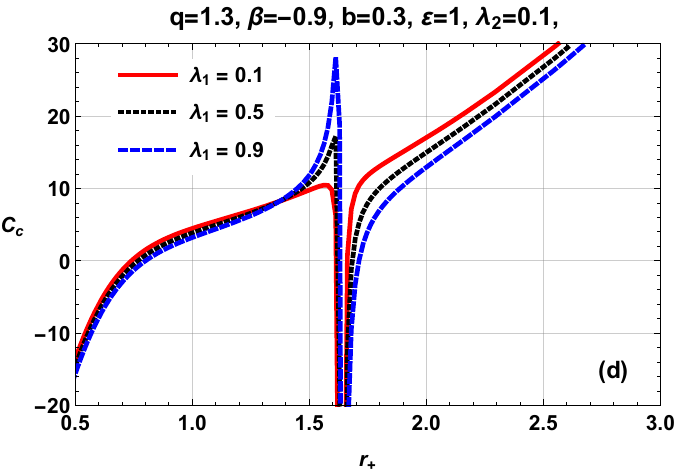}\\
       \vspace{0.4 cm}\includegraphics[scale=0.65]{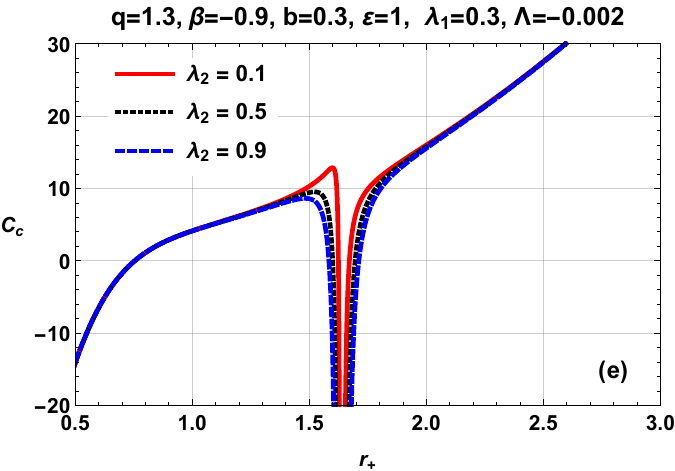}\hspace{5mm}
       \includegraphics[scale=0.65]{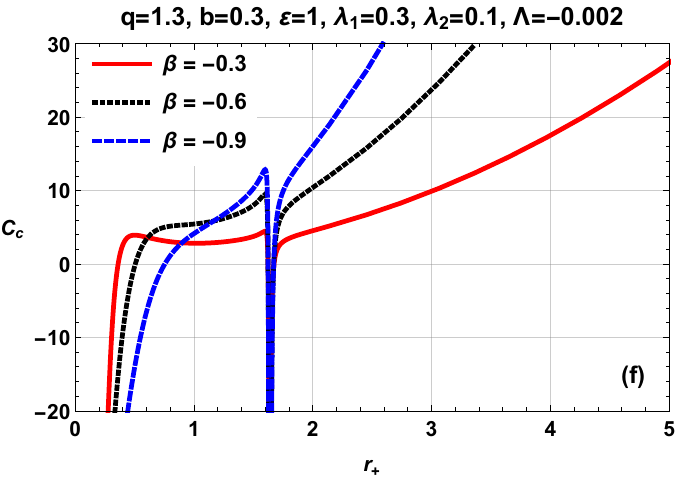}
       }
      	\caption{Variation of the black hole $C_c$ with the horizon radius $r_+$ using $\Lambda=-0.002$. }
      	\label{figCc01}
      \end{figure}
The specific heat of a black hole system is closely related to its thermodynamic stability. When specific heat is positive ($C > 0$), the system is stable, whereas a negative specific heat ($C<0$) indicates an unstable thermodynamic state. 
The corrected specific heat of the black hole system in usual notation is given as 
\begin{equation}
C_{c}=T_{H}\bigg(\frac{dS_{c}}{dT_{H}}\bigg).\label{cc1}
\end{equation}
For the current case of the H-AdS black hole system, Eq.(\ref{cc1}) with the help of Eq. (\ref{sc}), becomes
\begin{equation}
    C_c = \frac{2 \left[-\pi  \lambda _1 r_+ \left(r_+^3-2 q^3\right) \left(\text{cc}_1 \varepsilon  \left(\frac{b}{r_+}\right){}^{2/\beta }+(\beta -2) \beta \,  \text{cc}_2\right)+\beta \, \text{cc}_3 \lambda _2 r_+^2 \left(q^3+r_+^3\right){}^2+\pi ^2 \beta  \text{cc}_0 \left(2 q^6+q^3 r_+^3-r_+^6\right){}^2\right]}{\pi  r_+ \left(r_+^3-2 q^3\right){}^2 \left[(\beta -2) \beta \,  \text{cc}_5-\text{cc}_4 \varepsilon  \left(\frac{b}{r_+}\right){}^{2/\beta }\right]}.\label{C_c}
\end{equation}
The various terms involved in Eq. (\ref{C_c}) in the usual algebraic symbolism are cast as
{\begin{center}
$\text{cc}_0=\varepsilon  \left(\frac{b}{r_+}\right){}^{2/\beta } \left(2 (\beta +1) q^3-(\beta -2) r_+^3\right)+(\beta -2) \left(2 q^3+\Lambda  r_+^5-r_+^3\right),$ \\ $\text{cc}_1=2 (\beta +1) (3 \beta +4) q^9+3 (\beta  (7 \beta +4)+4) q^6 r_+^3-12 \beta ^2 q^3 r_+^6+2 (\beta -2) r_+^9$, \\ $\text{cc}_2=6 q^9-7 q^6 r_+^3 \left(\Lambda  r_+^2-3\right)+4 q^3 r_+^6 \left(\Lambda  r_+^2-3\right)+2 \Lambda  r_+^{11}$, \\ $\text{cc}_3=\varepsilon  \left(\frac{b}{r_+}\right){}^{2/\beta } \left((\beta -2) r_+^3-2 (\beta +1) q^3\right)-(\beta -2) \left(2 q^3+\Lambda  r_+^5-r_+^3\right)$, \\ $\text{cc}_4=2 (\beta +1) (\beta +2) q^6+2 (\beta  (5 \beta +3)+4) q^3 r_+^3-\left(\beta ^2-4\right) r_+^6$, \\ $\text{cc}_5=-2 q^6+2 q^3 r_+^3 \left(2 \Lambda  r_+^2-5\right)+\Lambda  r_+^8+r_+^6.$\end{center}}

For the detailed expression of uncorrected specific heat, one may look into the Section \ref{appen}.
A numerical platform is accordingly constructed to obtain Fig.\ref{figCc01}, illustrating the corrected specific heat $C_c$ (from Eq. \ref{C_c}) as an explicit function of $r_+$, with variation in $ q $, $ \varepsilon $, $ b $, $ \lambda_1 $, $ \lambda_2 $, and $ \beta $, keeping other parameters fixed (as indicated). It offers some valuable insights into the thermodynamic stability and phase transitions of a black hole system modified by a string fluid, Hayward regularisation, and quantum corrections discussed as follows. 

In Fig.\ref{figCc01}(a), due to the variation of $q$ the specific heat transitions from a negative to a positive value at a critical radius $r_c$, indicating a phase transition from an unstable small black hole (SBH) to a stable large black hole (LBH). A higher $q$-value lowers $r_c$, increasing instability in the SBH phase and enhancing stability in the LBH phase due to the smoothing effect of regularisation on singularities. Fig.\ref{figCc01}(b), varying $\varepsilon$ between $1$ and $-1$, shows a similar transition.  $\varepsilon = -1$ makes the SBH phase more unstable and the LBH phase more stable. It reflects the stabilising role of negative energy density. This feature is often associated with exotic matter in string theory. In Fig.\ref{figCc01}(c), the variation of $ b $ mirrors this pattern. Increasing $ b $ lowers $ r_c $ and destabilises the SBH phase and stabilises the LBH phase, suggesting that a stronger string fluid enhances stability at larger radii by increasing the effective energy density. Fig.\ref{figCc01}(d) shows that larger $ \lambda_1 $ values lower the critical radius $r_c$, increase SBH instability, and enhance LBH stability. This indicates that quantum corrections, possibly logarithmic in nature, refine the thermodynamic behaviour by favouring larger black holes. Fig.\ref{figCc01}(e) exhibits a similar trend—increasing $ \lambda_2 $ stabilises the LBH phase, likely due to geometric corrections affecting the horizon thermodynamics. Finally, Fig.\ref{figCc01}(f) shows that more negative $ \beta $ -values lower $ r_c $, destabilise the SBH phase and stabilise the LBH phase, reflecting how the non-standard metric regularisation favours larger, more stable black holes. This is a feature consistent with quantum gravity theories aimed at resolving singularities in GR.

\begin{figure}[t!]
      	\centering{
      	\includegraphics[scale=0.65]{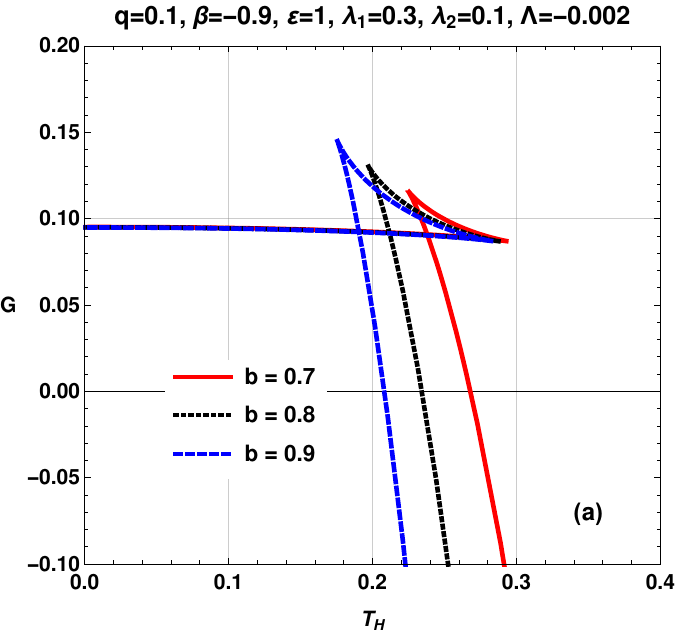} \hspace{5mm}
            \includegraphics[scale=0.65]{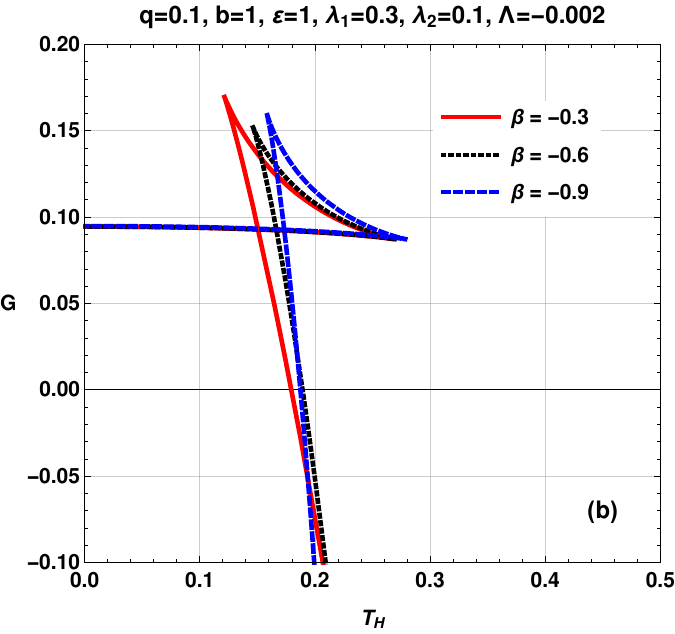}\\
            \vspace{0.4cm}
            \includegraphics[scale=0.65]{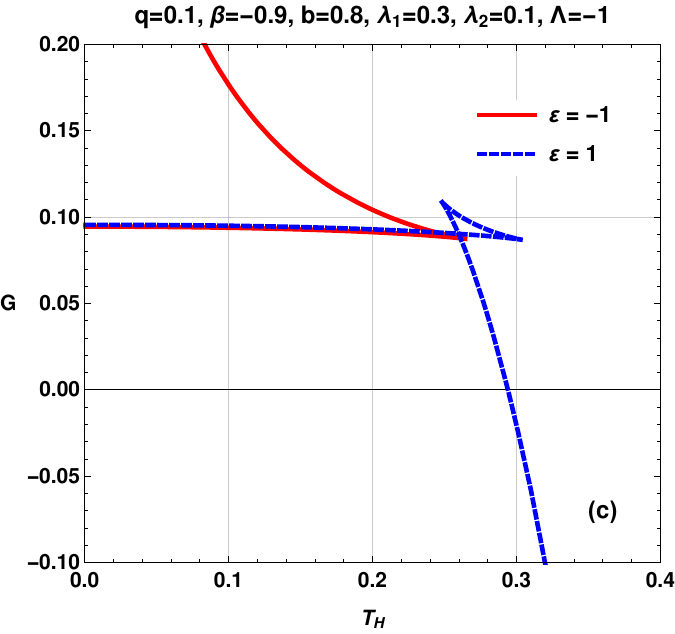}\hspace{5mm}
            \includegraphics[scale=0.65]{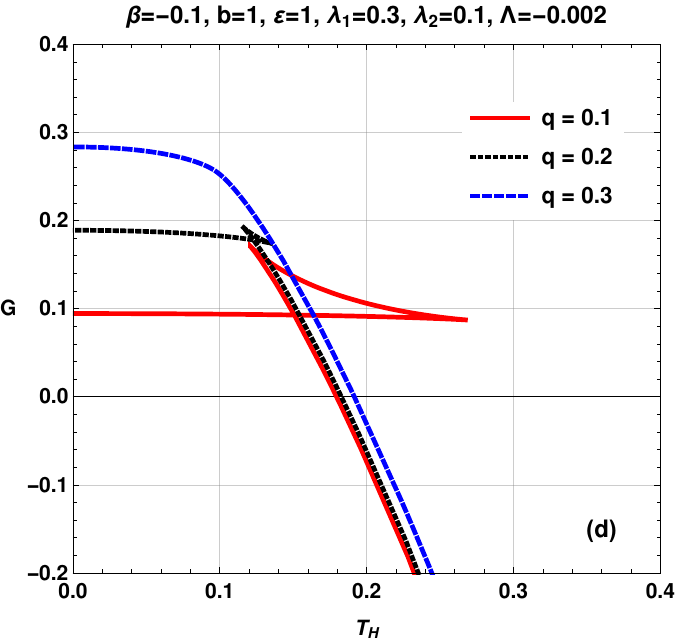}
       }
      	\caption{Variation of uncorrected Gibbs free energy ($G$) with Hawking temperature ($T_H$) for different black hole parametric conditions}
      	\label{figGvsT}
      \end{figure}
The Gibbs free energy of a system tells about the  stability of thermodynamic phases: the phase with the lowest $G$ (or $G_c$) is thermodynamically preferred. Phase transitions occur when $G$ (or $G_c$) changes discontinuously or when two phases have equal $G$, marking a critical temperature $T_c$. In the four subfigures of Fig.\ref{figGvsT} (using Eq.(\ref{G_unc})) we have plotted  $G$ against $T_H$ for different parameter variations keeping the values of other parameters as specified in the labels: $\beta = -0.9$, $\varepsilon = 1$, $q = 0.1$, $\Lambda = -0.002$ for most cases, with variations in specific subfigures. In Fig.\ref{figGvsT}(a), three curves are shown for $b = 0.7$, $b = 0.8$, and $b = 0.9$. The $G$ - $T_H$ plots exhibit a characteristic swallowtail shape, typical of black hole thermodynamics in extended phase spaces.
The swallowtail shape is a hallmark of first-order phase transitions in black hole thermodynamics, analogous to the van der Waals liquid-gas transition. The three branches of the swallowtail correspond to different black hole phases:\begin{itemize}
\item Small black hole (SBH) branch: The upper branch at low $T_H$, where $G$ decreases with increasing $T_H$. This phase has a smaller horizon radius $r_+$ and is typically unstable due to negative specific heat.
 \item Large black hole (LBH) branch: The lower branch at higher $T_H$, where $G$ continues to decrease with increasing $T_H$. This phase has a larger $r_+$ and is thermodynamically stable.
 \item Intermediate branch: The middle branch, which is unstable and connects the SBH and LBH branches, representing a metastable state.
 \end{itemize}
  
The intersection of the SBH and LBH branches (where $G$ is equal for both phases) marks the critical temperature $T_c \approx 0.25$, where a first-order phase transition occurs from a small to a large black hole. At this point, the black hole jumps from a smaller radius to a larger radius, releasing latent heat.
The parameter $b$ controls the strength of the string fluid, which contributes to the energy density of the system. As $b$ increases from 0.7 to 0.9, the critical temperature $T_c$ shifts slightly to lower values. This suggests that a stronger string fluid (higher $b$) facilitates the phase transition at a lower temperature, possibly by increasing the effective energy density, which mimics a pressure-like effect. The depth of the swallowtail (the maximum $G$ value at the peak) increases with $b$. A deeper swallowtail indicates a larger free energy difference between the phases, implying a more pronounced phase transition.
The $G$ values for the LBH branch become more negative with increasing $b$, indicating that the large black hole phase becomes more thermodynamically favoured as the string fluid strength increases.
For $T_H < T_c$, the SBH phase has a higher $G$ and is less stable, while for $T_H > T_c$, the LBH phase has a lower $G$ and is more stable. At $T_H = T_c$, the two phases coexist, and the system can transition between them without a change in $G$, consistent with a first-order phase transition.
 The string fluid, parameterised by $b$, introduces a non-standard matter component to the black hole system. In the context of string theory, such fluids can arise from the presence of fundamental strings or branes, which modify the spacetime geometry and thermodynamics. The negative $\beta = -0.9$ suggests a non-standard regularity condition, related to a modified black hole metric that avoids singularities, such as in the Hayward regularisation context.

As in Fig.\ref{figGvsT}(b), we depict three distinct G-curves for $\beta = -0.3$, $\beta = -0.6$, and $\beta = -0.9$. It also exhibits a swallowtail shape. As mentioned before, the swallowtail shape indicates a first-order phase transition between small and large black hole phases. The critical temperature $T_c \approx 0.2$ is consistent across the curves, suggesting that $\beta$ has a weaker influence on $T_c$ compared to $b$.
The parameter $\beta$ is a generalisation parameter affecting the regularity of the black hole system. As $\beta$ becomes more negative (from -0.3 to -0.9), the depth of the swallowtail increases slightly, indicating a more pronounced phase transition. This suggests that a more negative $\beta$ enhances the free energy difference between the SBH and LBH phases.
The $G$ values for the LBH branch become slightly more negative with a more negative $\beta$, implying that the large black hole phase becomes more stable as $\beta$ decreases.The SBH branch at low $T_H$ shows a steeper rise for a more negative $\beta$, indicating that the small black hole phase becomes less stable, possibly due to a stronger modification of the metric near the horizon.
The negative $\beta$ values suggest a non-standard black hole metric, possibly one designed to resolve singularities (e.g., Hayward or Bardeen-like metrics). Such metrices often introduce a minimum length scale (related to $q$), which modifies the thermodynamic potentials. A more negative $\beta$ may enhance this regularisation effect, leading to a more stable large black hole phase by smoothing out singularities that would otherwise destabilise the system.
In the context of modified gravity or quantum gravity theories, $\beta$ could be related to a deformation parameter that alters the spacetime geometry. The negative values might correspond to a regime where the black hole metric deviates significantly from the Schwarzschild or Reissner-Nordstr\"{o}m solutions, potentially incorporating quantum or stringy effects as stabilising influences at larger event horizon radii.

As in Fig.\ref{figGvsT}(c), variation of $G$ versus $T_H$ is plotted for two different values of $\varepsilon$. Here also, for positive values of the model parameter, the curves have the characteristic swallowtail structure, confirming the presence of a first-order phase transition between small and large black hole phases. But when $\varepsilon$ has a negative value, the phase transition disappears. The figure indicates that the sign of energy density has a noticeable impact on the phase transition behaviour of the black hole. Fig.\ref{figGvsT}(d), shows three distinct curves for $q = 0.1$, $q = 0.2$, and $q = 0.3$. The $G$-$T_H$ plots again exhibit a swallowtail shape.
As $q$ increases from $0.1$ to $0.3$, the critical temperature $T_c$ shifts to lower values. This suggests that a larger regularisation scale facilitates the phase transition at a lower temperature. The depth of the swallowtail increases with $q$, indicating a more pronounced phase transition. The $G$ values for the LBH branch become more negative with increasing $q$, implying that the large black hole phase becomes more stable as the regularisation scale increases. The SBH branch shows a steeper rise for larger $q$, indicating that the small black hole phase becomes less stable.

The Hayward regularisation scale $q$ introduces a minimum length scale to the black hole metric, avoiding singularities at the origin. This is similar to the effect of charge in Reissner-Nordstr\"{o}m black holes, where the charge $Q$ (analogous to $q$) modifies the geometry and thermodynamics. A larger $q$ effectively increases the ``charge-like" contribution, which lowers the critical temperature and stabilises the large black hole phase by reducing the curvature near the horizon.
 The Hayward regularisation is a phenomenological approach to resolving singularities in black hole spacetimes, often motivated by quantum gravity or string theory. The parameter $q$ acts as a proxy for quantum effects that smooth out the singularity, and its increase leads to a more stable thermodynamic system, as seen in the preference for the LBH phase.
 
\begin{figure}[t!]
      	\centering{
    \includegraphics[scale=0.65]{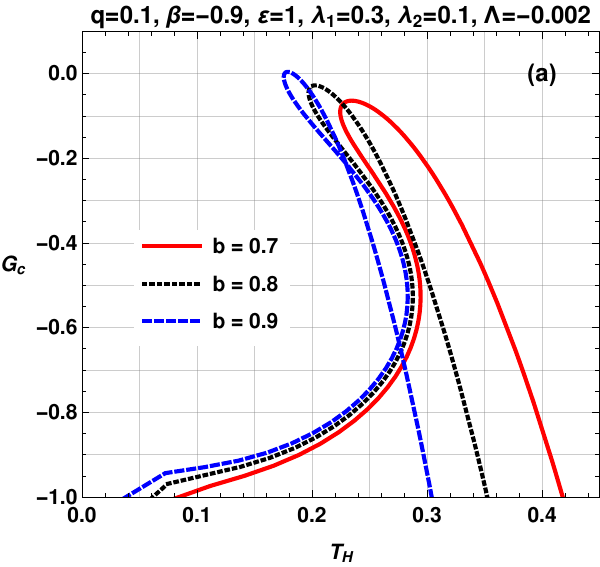}\hspace{5mm}
    \includegraphics[scale=0.65]{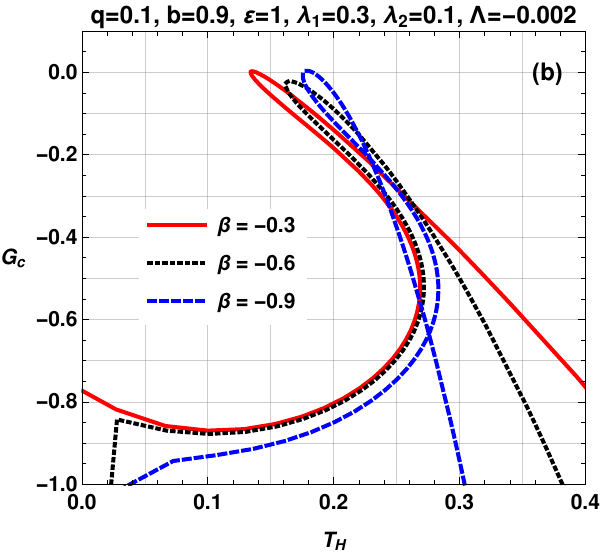}\\ \vspace{4mm} \includegraphics[scale=0.65]{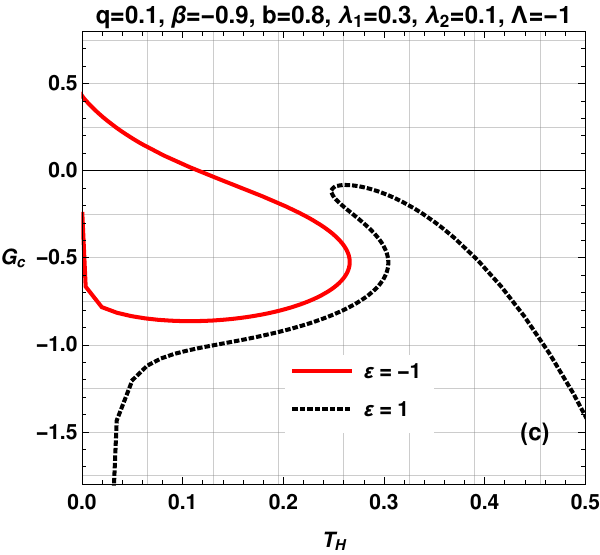}\hspace{5mm}
    \includegraphics[scale=0.65]{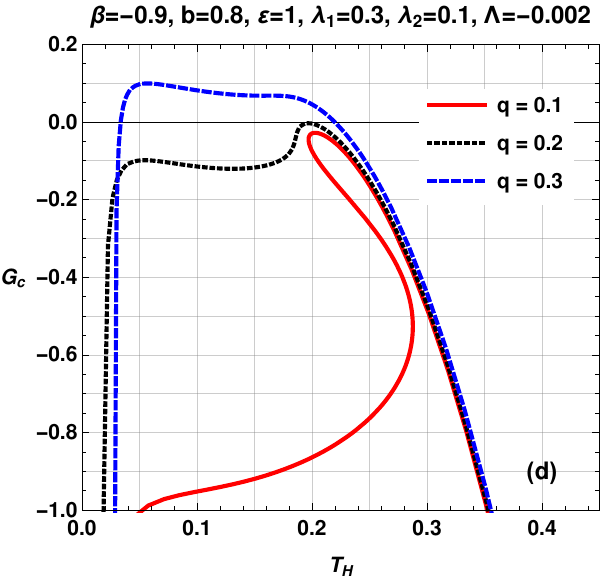}\\
    \vspace{4mm}\includegraphics[scale=0.65]{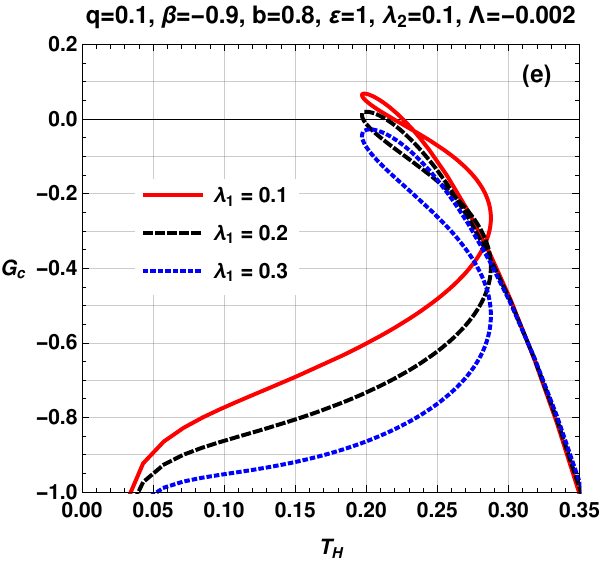}\hspace{5mm}
    \includegraphics[scale=0.65]{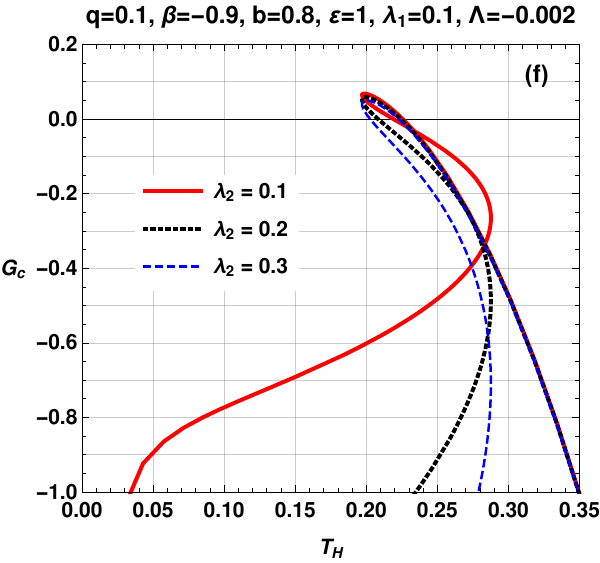}}
    \caption{Variation of corrected Gibbs free energy $G_c$ with Hawking temperature $T_H$}
    \label{figGcvsT}
    \end{figure}
As a graphical representation of Eq.(\ref{Gc}), we have shown the corrected Gibbs free energy $G_c$ vs $T_H$ in Fig.\ref{figGcvsT}. This figure contains six subfigures, each plotting $G_c$ against $T_H$ for different parameter variations, with fixed values of other parameters as specified in the labels. 
Across all six subfigures, the swallowtail structure observed in the $G_c$ vs. $T_H$ plots signifies the presence of a first-order phase transition between small black hole (SBH) and large black hole (LBH) phases, occurring consistently near the critical temperature $T_c \approx 0.25$. The introduction of corrected entropy $S_c$, incorporating logarithmic and geometric terms via the correction parameters $\lambda_1$ and $\lambda_2$, refines this transition, smoothing the swallowtail and subtly shifting the thermodynamic behaviour. In Fig.\ref{figGcvsT}(a), increasing the parameter $b$ lowers $T_c$, deepens the swallowtail, and stabilizes the LBH phase, indicating that higher values of $b$ enhance the thermodynamic favorability of the large black hole. Fig.\ref{figGcvsT}(b) shows a similar trend with varying $\beta$: more negative values increase the depth of the swallowtail, suggesting a stronger first-order transition and greater LBH stability, especially when entropy corrections are included. In Fig.\ref{figGcvsT}(c), the effect of $\varepsilon$ is clearly evident. For $\varepsilon=+1$, the swallowtail structure is seen, which confirms the first-order phase transition between SBH and LBH phase. However, this structure becomes deformed for $\varepsilon=-1$. This suggests that the phase transition is stronger when the string fluid has positive energy density. Fig.\ref{figGcvsT}(d) illustrates how increasing the Hayward regularisation parameter $q$ similarly lowers $T_c$ and stabilises the LBH phase, especially when quantum corrections are prominent. Fig.\ref{figGcvsT}(e) and Fig.\ref{figGcvsT}(f) focus directly on the roles of $\lambda_1$ and $\lambda_2$ respectively, showing that increasing either correction parameter deepens the swallowtail and enhances the thermodynamic favourability of the LBH phase while making the SBH phase less stable. These behaviours collectively indicate that both classical parameters (such as $b, \beta, \varepsilon, q$) and quantum corrections (through $\lambda_1, \lambda_2$) interplay to govern the phase structure, with higher-order corrections-arising possibly from loop quantum gravity, string theory, or thermal fluctuations-playing a significant role in refining the nature of black hole phase transitions in this framework.

The corrected critical temperature corresponding to $G_c=0$ is given by
\begin{equation}
    T^{min}_c(corrected) = -\frac{\left(q^3+r_+^3\right) \left((\beta -2) \left(\Lambda  r_+^2-3\right)-3 \beta  \varepsilon  \left(\frac{b}{r_+}\right){}^{2/\beta }\right)}{6 (\beta -2) r_+^2 \left(-\lambda _1 \log \left(A_1\right)+\lambda _1 \log (16 \pi )+\frac{\lambda _2 r_+}{\pi  \left(r_+^3-2 q^3\right)}+\pi  \left(r_+^2-\frac{2 q^3}{r_+}\right)\right)},
\end{equation}
where,
\begin{equation}
A_1 = \frac{\left(r_+^3-2 q^3\right) \left(\varepsilon  \left(\frac{b}{r_+}\right){}^{2/\beta } \left(2 (\beta +1) q^3-(\beta -2) r_+^3\right)+(\beta -2) \left(2 q^3+\Lambda  r_+^5-r_+^3\right)\right){}^2}{(\beta -2)^2 r_+^3 \left(q^3+r_+^3\right){}^2}.\end{equation}
The uncorrected critical temperature is 
\begin{equation}
    T^{min}_c = -\frac{\left(q^3+r_+^3\right) \left(\beta  \left(-3 \varepsilon  \left(\frac{b}{r_+}\right){}^{2/\beta }+\Lambda  r_+^2-3\right)-2 \Lambda  r_+^2+6\right)}{6 \pi  (\beta -2) r_+ \left(r_+^3-2 q^3\right)}.
\end{equation}

\section{Results and Discussion}\label{sec7}

The analysis of the thermodynamic behaviours of a black hole modified by a string fluid, Hayward regularisation, and quantum corrections reveals a rich and intricate structure that significantly departs from classical black hole thermodynamics. By essentially relaxing its thermodynamic behaviour, a string fluid modifies the phase structure of a Hayward-AdS black hole.  The string density introduces an additional term into the metric, which shifts the associated Gibbs energy ``swallowtail" to lower temperatures and lowers the critical temperature and pressure at which the small-large black hole transition takes place. If the tension of the string fluid is high enough, it can erase the phase transition and behave as an additional matter component that weakens it. The presence of first-order phase transitions, modulated by several physical parameters, and the emergence of stable large black hole (LBH) phases under various modifications illustrate the collective complex interplay among classical gravity, exotic matter fields, and quantum corrections simultaneously.

We have seen that the parameters $q,\beta, b$ and $\varepsilon$ and the small correction terms $\lambda_1$ and $\lambda_2$ primarily govern the thermodynamic behaviour of the system. The role of each parameter in shaping the phase structure provides valuable physical insights. The string fluid strength parameter $b$ and the energy density parameter $\varepsilon$ appear as effective matter couplings, with increasing $b$ and negative $\varepsilon$ both contributing to the stabilization of the LBH phase. The thermodynamic effect of the string fluid can be interpreted as an increase in the effective energy density surrounding the black hole, thereby altering the curvature and pressure terms in the gravitational field equations. This effect closely mimics the inclusion of additional degrees of freedom, which in thermodynamic systems generally promote phase stability and higher entropy states. Notably, the stabilising influence of a negative energy density though - theoretically exotic - is consistent with predictions from certain effective field theories and cosmological models, particularly those invoking dark energy or phantom matter components. $\varepsilon$ expresses whether the surrounding string fluid pushes or pulls on spacetime and how dense or tensile it is. Similar to how surface tension smooths out ripples on a liquid surface, increasing $b$ flattens out abrupt fluctuations in temperature, enthalpy, and heat capacity.  Like external pressure can move a liquid's boiling point, it also causes phase transitions to shift to somewhat larger horizon radii. Hence, we can consider $b$ to be analogous to an external pressure or stabilising tension that is supplied to the system; it makes the thermodynamics of the black hole more stable and gentle, particularly for small black holes.

The Hayward regularisation scale $q$ further enhances stability and lowers the critical temperature $T_c$. This is significant because it reaffirms the physical interpretation of regular black holes as avoiding the pathological features of classical singularities. In the Hayward model, $q$ effectively acts as a nonlinear electrodynamic charge or a parameter encoding quantum geometric effects near the core. $q$ alters temperature, entropy, and stability by an internal repulsion, much like electric charge does in Reissner-Nordstr\"{o}m black holes. It increases the horizon size at which characteristic features emerge: it increases the corrected entropy $S_c$, lifts the enthalpy $H_c$, shifts the Helmholtz free energy minimum $F_c$, moves the divergence of the heat capacity $C_c$ to larger radii, and raises the Hawking temperature $T_H$ for small black holes, consequently broadening the stable phase. Its thermodynamic influence, analogous to the role of electric charge $Q$ in Reissner-Nordstr\"{o}m-AdS black holes, is to introduce repulsive effects that modify the pressure-volume structure of the black hole interior. This mechanism naturally leads to a lower energy threshold for phase transitions and a more stable end state in the form of an LBH.

The behaviour close to the origin is softened by the parameter $\beta$; more negative values shift the extrema of $H_c$ and $F_c$ outward, relocate the pole of $C_c$ to a bigger $r_+$, and lower $T_H$ and $S_c$ at small radii while augmenting them at large radii, favouring stability of larger black holes. Phase transitions are pushed outward by the parameter $b$, which also smooths the slope of $S_c$, lowers $H_c$ near the origin, lowers $F_c$ at small $r_+$, and regularises $C_c$. 

Quantum corrections, parameterised by $\lambda_1$ and $\lambda_2$, are pivotal in smoothing the phase transition and enhancing LBH stability. These parameters are associated with logarithmic or higher-order corrections to the Bekenstein-Hawking entropy, derived from statistical fluctuations, loop quantum gravity, or string-theoretic considerations. The key effect of these corrections is the modification of the entropy-area relation, which in turn adjusts the temperature and free energy profiles. As a result, the system exhibits a more continuous thermodynamic evolution, with reduced divergence at $T_c$ and enhanced robustness of the LBH phase. This result supports a growing body of evidence suggesting that quantum gravitational effects, far from being negligible, play a crucial role in determining the endpoint of black hole evolution and in resolving classical instabilities.

The results obtained in this study establish some interesting similarities between first-order phase transition in the H-AdS system and the liquid-gas phase transition in van der Waals fluids. This analogy is substantiated by the characteristic swallowtail structure observed in the Gibbs free energy $G$ and corrected Gibbs free energy $G_c$ versus Hawking temperature $T_H$ diagrams. These swallowtails describe the coexisting regions of SBH and LBH phases, with the intersection point defining the critical temperature $T_c$. Such a thermodynamic signature indicates the existence of systems exhibiting latent heat and discontinuous changes in thermodynamic state variables - a phenomena now shown to be plausible in black holes subject to quantum and string-theoretic refining modifications.

A particularly noteworthy aspect of these findings is the synergistic action of all modifications-string fluid, regularisation, and quantum corrections- in achieving thermodynamic stability. While classical black holes, such as the Schwarzschild solution, are thermodynamically unstable due to their negative specific heat, the inclusion of the above mechanisms systematically removes this instability. The SBH phase remains unstable (with $C < 0$), consistent with its association to evaporation and high curvature regions, but the LBH phase attains positive specific heat ($C > 0$), indicating the capacity for stable equilibrium with a surrounding heat bath. This result not only reinforces the physical plausibility of non-singular, stable black hole configurations but also presents a compelling model for black hole remnants or long-lived equilibrium states, potentially observable in diverse high-energy astrophysical phenomena.

From a theoretical perspective, these results provide a controlled setting to explore the convergence of classical and quantum gravitational physics. The inclusion of the Hayward regularisation and quantum entropy corrections is motivated by approaches such as loop quantum gravity and non-singular cosmologies, where geometric quantisation leads to modifications of the Einstein equations at small scales. Similarly, the incorporation of a string fluid reflects effective descriptions of brane or string networks in low-energy limits of string theory, with implications for holography and AdS/CFT duality. The observed phase structure thus mirrors, at least qualitatively, the thermodynamics of gauge theories with confinement-deconfinement transitions, further reinforcing the deep connections between gravity, thermodynamics, and quantum field theory.

The system's similarity to Reissner-Nordstr\"{o}m-AdS black holes extends beyond superficial analogies. In both cases, the existence of a critical point, the swallowtail structure of $G$ vs $T$, and the discontinuity in specific heat suggest that the black hole behaves as a thermodynamic fluid obeying a generalized equation of state. The identification of parameters such as $q$ and $b$ as analogues to charge and interaction strength provides a new lens through which to interpret modified gravity theories. The presence of a small negative cosmological constant ($\Lambda = -0.002$), while subdominant in the present case, subtly contributes to the thermodynamic behaviour, embedding the system within a de Sitter-like or asymptotically quasi-flat background. This contextual shift enables broader interpretations within the framework of extended thermodynamics, where pressure-volume terms are introduced and the cosmological constant is treated as a thermodynamic variable.

On the observational side, the study invites further exploration into the possible astrophysical signatures of such phase transitions. While direct detection of thermodynamic quantities like specific heat is beyond current observational capability, indirect signatures may manifest in black hole shadow profiles, quasinormal mode spectra, or gravitational wave emissions during mergers. For instance, the presence of a stable LBH phase implies the existence of equilibrium configurations that may differ from classical expectations, potentially affecting the ringdown phase of black hole mergers. Similarly, quantum corrections to entropy could influence the black hole area spectrum, which might be detectable in high-precision measurements of Hawking radiation spectra or via deviations in entropy bounds during accretion processes.
An intriguing implication of this work lies in its potential application to black hole evaporation and the information paradox. The stability of the LBH phase suggests the possibility of black hole remnants - end states of evaporation that retain finite mass and entropy. Such remnants could evade the singularity and information loss problems by maintaining a unitary evolution throughout the black hole lifecycle. If supported by further theoretical and observational studies, this scenario could provide a natural resolution to long-standing paradoxes in black hole physics.

\section{Concluding Remarks}\label{sec8}
The thermodynamic framework developed here offers fertile ground for future generalisations in the context of H-AdS black hole thermodynamics. One promising direction is the inclusion of rotation or angular momentum, which would extend the analysis to Kerr-like black holes with modified geometries. Another extension involves coupling to other exotic matter fields, such as phantom energy or non-minimally coupled scalar fields, which could further enrich the phase structure and stability criteria. Moreover, the study of dynamic processes - such as quench-induced transitions or thermal cycling - within this thermodynamic setting could unveil non-equilibrium phenomena, offering a more integrated picture of black hole evolution in diverse realistic astro-cosmic environments.
Additionally, from the standpoint of quantum gravity phenomenology, the results presented here emphasize the importance of combining multiple theoretical modifications to achieve physically meaningful models. No single parameter-be it $b$, $q$, or $\lambda_i$-is solely responsible for the observed stabilisation; rather, it is their collective interaction that gives rise to a well-behaved thermodynamic system. This underscores the necessity of a holistic approach in constructing viable black hole models that account for both classical gravitational dynamics and quantum corrections. In doing so, such models not only bridge the gap between general relativity and quantum field theory but also provide testable predictions that could shape our understanding of black holes, both as fundamental objects and as astrophysical phenomena.
In conclusion, the present study establishes a comprehensive and physically consistent picture of black hole thermodynamics in a framework augmented by string fluids, Hayward regularization, and quantum corrections. The emergence of first-order phase transitions, stabilisation of the LBH phase, and modulation of critical behaviour by tunable parameters reflect the depth and versatility of the model. These findings reinforce the notion that black holes are thermodynamic systems governed by principles analogous to those of conventional matter, while also illuminating the profound effects of quantum and string-theoretic physics in their structure and evolution. As observational capabilities advance and theoretical tools become more refined, the predictions of such models may soon be subjected to empirical scrutiny, offering a unique window into the interface of gravity, quantum theory, and high-energy astrophysics.
\section{Appendix: thermo-parametric analysis} \label{appen}
\begin{itemize}
    \item{\bf Corrected enthalpy}: The expression for corrected enthalpy, as derived from Eq.(\ref{Enthalpy}) is explicitly given as\\ \vspace{-1cm}
    \begin{equation}
\begin{aligned}
   \hspace{-1 cm} H_c =& \frac{1}{4 \pi  (\beta -2)}\Bigg[\frac{2 \pi  \beta ^2 \varepsilon  r_+ \left(\frac{b}{r_+}\right){}^{2/\beta }}{\beta -2}+\frac{3 \beta ^2 \varepsilon  \, _2F_1\left(1,\frac{2 (\beta -1)}{3 \beta };\frac{5}{3}-\frac{2}{3 \beta };-\frac{r_+^3}{q^3}\right) r_+^2 \lambda _1 \left(\frac{b}{r_+}\right){}^{2/\beta }}{q^3 (\beta -1)}  \\&+\frac{6 \beta  \varepsilon  \, _2F_1\left(1,\frac{2 (\beta -1)}{3 \beta };\frac{5}{3}-\frac{2}{3 \beta };-\frac{r_+^3}{q^3}\right) r_+^2 \lambda _1 \left(\frac{b}{r_+}\right){}^{2/\beta }}{q^3 (\beta -1)}+\frac{6 \beta ^2 \varepsilon  \lambda _1 \left(\frac{b}{r_+}\right){}^{2/\beta }}{\beta  r_++2 r_+}+\frac{14 \beta  \varepsilon  \lambda _1 \left(\frac{b}{r_+}\right){}^{2/\beta }}{\beta  r_++2 r_+}\\& +\frac{8 \varepsilon  \lambda _1 \left(\frac{b}{r_+}\right){}^{2/\beta }}{\beta  r_++2 r_+}+\frac{\beta ^2 \varepsilon  \, _2F_1\left(1,1-\frac{2}{3 \beta };2-\frac{2}{3 \beta };\frac{r_+^3}{2 q^3}\right) r_+^3 \lambda _2 \left(\frac{b}{r_+}\right){}^{2/\beta }}{2 \pi  q^6 (3 \beta -2)}+\frac{\beta  \varepsilon  \, _2F_1\left(1,1-\frac{2}{3 \beta };2-\frac{2}{3 \beta };\frac{r_+^3}{2 q^3}\right) r_+^3 \lambda _2 \left(\frac{b}{r_+}\right){}^{2/\beta }}{2 \pi  q^6 (3 \beta -2)}\\&+\frac{3 \beta  \varepsilon  \, _2F_1\left(2,1-\frac{2}{3 \beta };2-\frac{2}{3 \beta };\frac{r_+^3}{2 q^3}\right) r_+^3 \lambda _2 \left(\frac{b}{r_+}\right){}^{2/\beta }}{2 \pi  q^6 (3 \beta -2)}-\frac{\beta ^2 \varepsilon  \lambda _2 \left(\frac{b}{r_+}\right){}^{2/\beta }}{2 \pi  q^3}-\frac{\beta  \varepsilon  \lambda _2 \left(\frac{b}{r_+}\right){}^{2/\beta }}{2 \pi  q^3}-\frac{4 \pi  \beta  \varepsilon  r_+ \left(\frac{b}{r_+}\right){}^{2/\beta }}{\beta -2}\\&-\frac{9 \beta ^2 \varepsilon  \, _2F_1\left(2,\frac{2 (\beta -1)}{3 \beta };\frac{5}{3}-\frac{2}{3 \beta };-\frac{r_+^3}{q^3}\right) r_+^2 \lambda _1 \left(\frac{b}{r_+}\right){}^{2/\beta }}{q^3 (\beta -1)}-\frac{3 \beta  \varepsilon  \, _2F_1\left(1,\frac{2 (\beta -1)}{3 \beta };\frac{5}{3}-\frac{2}{3 \beta };\frac{r_+^3}{2 q^3}\right) r_+^2 \lambda _1 \left(\frac{b}{r_+}\right){}^{2/\beta }}{2 q^3 (\beta -1)}\\&+\frac{2 \pi  q^3 \beta ^2 \varepsilon  \left(\frac{b}{r_+}\right){}^{2/\beta }}{(\beta +1) r_+^2}+\frac{2 \pi  q^3 \beta  \varepsilon  \left(\frac{b}{r_+}\right){}^{2/\beta }}{(\beta +1) r_+^2}-\frac{2}{3} \pi  (\beta -2) \Lambda  r_+^3+\frac{6 (\beta -2) \lambda _1}{r_+}+2 (\beta -2) r_+ \left(2 \Lambda  \lambda _1+\pi \right)\\&+\frac{(\beta -2) \log \left(r_+\right) \lambda _2}{\pi  q^3}+\frac{2 (\beta -2) \Lambda  r_+^2 \lambda _2}{\pi  \left(6 q^3-3 r_+^3\right)}+\frac{2^{2/3} (\beta -2) \Lambda  \log \left(2 q-2^{2/3} r_+\right) \left(3\ 2^{2/3} \pi  \lambda _1 q^2+2 \lambda _2\right)}{9 \pi  q}\\&-\frac{(\beta -2) \Lambda  \log \left(2 q^2+2^{2/3} r_+ q+\sqrt[3]{2} r_+^2\right) \left(3\ 2^{2/3} \pi  \lambda _1 q^2+2 \lambda _2\right)}{9 \sqrt[3]{2} \pi  q}-\frac{2^{2/3} (\beta -2) \Lambda  \tan ^{-1}\left(\frac{q+2^{2/3} r_+}{\sqrt{3} q}\right) \left(3\ 2^{2/3} \pi  q^2 \lambda _1-2 \lambda _2\right)}{3 \sqrt{3} \pi  q}\\&-\frac{(\beta -2) \log \left(r_+^3-2 q^3\right) \lambda _2}{3 \pi  q^3}+\frac{2 \pi  q^3 (\beta -2)}{r_+^2}-\frac{2 (\beta -2) r_+ \left(\Lambda  q^3+3 r_+\right) \lambda _1}{q^3+r_+^3}\Bigg]. \label{Hc}
\end{aligned}
\end{equation}
\end{itemize}
\begin{itemize}
    \item {\bf Corrected internal energy}:\\ \vspace{-1cm}
\begin{equation}
\begin{aligned}
     \hspace{-1 cm} U_c = &\frac{1}{24 (\beta -2)^2}\Bigg[12 \beta ^2 \varepsilon  r_+ \left(\frac{b}{r_+}\right){}^{2/\beta }-24 \beta  \varepsilon  r_+ \left(\frac{b}{r_+}\right){}^{2/\beta }+\frac{18 (\beta -2) \beta ^2 \varepsilon  \, _2F_1\left(1,\frac{2 (\beta -1)}{3 \beta };\frac{5}{3}-\frac{2}{3 \beta };-\frac{r_+^3}{q^3}\right) r_+^2 \lambda _1 \left(\frac{b}{r_+}\right){}^{2/\beta }}{\pi  q^3 (\beta -1)}\\& +\frac{36 (\beta -2) \beta  \varepsilon  \, _2F_1\left(1,\frac{2 (\beta -1)}{3 \beta };\frac{5}{3}-\frac{2}{3 \beta };-\frac{r_+^3}{q^3}\right) r_+^2 \lambda _1 \left(\frac{b}{r_+}\right){}^{2/\beta }}{\pi  q^3 (\beta -1)}+\frac{36 (\beta -2) \beta ^2 \varepsilon  \lambda _1 \left(\frac{b}{r_+}\right){}^{2/\beta }}{\pi  (\beta +2) r_+}+\frac{48 (\beta -2) \varepsilon  \lambda _1 \left(\frac{b}{r_+}\right){}^{2/\beta }}{\pi  (\beta +2) r_+}\\&+\frac{84 (\beta -2) \beta  \varepsilon  \lambda _1 \left(\frac{b}{r_+}\right){}^{2/\beta }}{\pi  (\beta +2) r_+}+\frac{3 (\beta -2) \beta ^2 \varepsilon  \, _2F_1\left(1,1-\frac{2}{3 \beta };2-\frac{2}{3 \beta };\frac{r_+^3}{2 q^3}\right) r_+^3 \lambda _2 \left(\frac{b}{r_+}\right){}^{2/\beta }}{\pi ^2 q^6 (3 \beta -2)}\\&+\frac{3 (\beta -2) \beta  \varepsilon  \, _2F_1\left(1,1-\frac{2}{3 \beta };2-\frac{2}{3 \beta };\frac{r_+^3}{2 q^3}\right) r_+^3 \lambda _2 \left(\frac{b}{r_+}\right){}^{2/\beta }}{\pi ^2 q^6 (3 \beta -2)}+\frac{9 (\beta -2) \beta  \varepsilon  \, _2F_1\left(2,1-\frac{2}{3 \beta };2-\frac{2}{3 \beta };\frac{r_+^3}{2 q^3}\right) r_+^3 \lambda _2 \left(\frac{b}{r_+}\right){}^{2/\beta }}{\pi ^2 q^6 (3 \beta -2)}\\&-\frac{3 (\beta -2) \beta ^2 \varepsilon  \lambda _2 \left(\frac{b}{r_+}\right){}^{2/\beta }}{\pi ^2 q^3}-\frac{3 (\beta -2) \beta  \varepsilon  \lambda _2 \left(\frac{b}{r_+}\right){}^{2/\beta }}{\pi ^2 q^3}-\frac{9 (\beta -2) \beta  \varepsilon  \, _2F_1\left(1,\frac{2 (\beta -1)}{3 \beta };\frac{5}{3}-\frac{2}{3 \beta };\frac{r_+^3}{2 q^3}\right) r_+^2 \lambda _1 \left(\frac{b}{r_+}\right){}^{2/\beta }}{\pi  q^3 (\beta -1)}\\&-\frac{54 (\beta -2) \beta ^2 \varepsilon  \, _2F_1\left(2,\frac{2 (\beta -1)}{3 \beta };\frac{5}{3}-\frac{2}{3 \beta };-\frac{r_+^3}{q^3}\right) r_+^2 \lambda _1 \left(\frac{b}{r_+}\right){}^{2/\beta }}{\pi  q^3 (\beta -1)}+\frac{12 q^3 (\beta -2) \beta ^2 \varepsilon  \left(\frac{b}{r_+}\right){}^{2/\beta }}{(\beta +1) r_+^2}+\frac{12 q^3 (\beta -2) \beta  \varepsilon  \left(\frac{b}{r_+}\right){}^{2/\beta }}{(\beta +1) r_+^2}\\&-24 (\beta -2) r_++12 (\beta -2) \beta  r_++\frac{36 (\beta -2) \beta  \lambda _1}{\pi  r_+}+\frac{6 (\beta -2) \beta  \log \left(r_+\right) \lambda _2}{\pi ^2 q^3}+\frac{4 (\beta -2) \log \left(r_+^3-2 q^3\right) \lambda _2}{\pi ^2 q^3}\\&-\frac{12 (\beta -2) \log \left(r_+\right) \lambda _2}{\pi ^2 q^3}-\frac{2 (\beta -2) \beta  \log \left(r_+^3-2 q^3\right) \lambda _2}{\pi ^2 q^3}-\frac{72 (\beta -2) \lambda _1}{\pi  r_+}-\frac{24 q^3 (\beta -2)}{r_+^2}+\frac{72 (\beta -2) r_+^2 \lambda _1}{\pi  \left(q^3+r_+^3\right)}\\& +\frac{12 q^3 (\beta -2) \beta }{r_+^2}-\frac{36 (\beta -2) \beta  r_+^2 \lambda _1}{\pi  \left(q^3+r_+^3\right)}\Bigg].\label{Uc}
\end{aligned}
\end{equation}
\end{itemize}
\begin{itemize}
    \item {\bf Helmholtz free energy}:
The explicit expression for corrected Helmholtz free energy given by Eq. \eqref{Fc} is
\begin{equation}
\begin{aligned}
         \hspace{-1cm}F_c =&\frac{1}{24 \pi ^2}\Bigg[6 \pi ^2 \beta ^2 \varepsilon  r_+ \left(\frac{b}{r_+}\right){}^{2/\beta }{(\beta -2)^2}+\frac{162 \pi ^2 \beta ^2 \varepsilon  \, _2F_1\left(2,\frac{\beta -2}{3 \beta };\frac{4}{3}-\frac{2}{3 \beta };-\frac{r_+^3}{q^3}\right) r_+ \left(\frac{b}{r_+}\right){}^{2/\beta }}{(\beta -2)^2}
         \\& +\frac{18 \pi  \beta ^2 \varepsilon  \, _2F_1\left(1,\frac{2 (\beta -1)}{3 \beta };\frac{5}{3}-\frac{2}{3 \beta };-\frac{r_+^3}{q^3}\right) r_+^2 \lambda _1 \left(\frac{b}{r_+}\right){}^{2/\beta }}{q^3 (\beta -2) (\beta -1)}+\frac{36 \pi  \beta  \varepsilon  \, _2F_1\left(1,\frac{2 (\beta -1)}{3 \beta };\frac{5}{3}-\frac{2}{3 \beta };-\frac{r_+^3}{q^3}\right) r_+^2 \lambda _1 \left(\frac{b}{r_+}\right){}^{2/\beta }}{q^3 (\beta -2) (\beta -1)}\\
        & +\frac{9 \pi  \beta ^2 \varepsilon  \, _2F_1\left(1,\frac{2 (\beta -1)}{3 \beta };\frac{5}{3}-\frac{2}{3 \beta };-\frac{r_+^3}{q^3}\right) \log (16 \pi ) r_+^2 \lambda _1 \left(\frac{b}{r_+}\right){}^{2/\beta }}{q^3 (\beta -2) (\beta -1)}\\& +\frac{18 \pi  \beta  \varepsilon  \, _2F_1\left(1,\frac{2 (\beta -1)}{3 \beta };\frac{5}{3}-\frac{2}{3 \beta };-\frac{r_+^3}{q^3}\right) \log (16 \pi ) r_+^2 \lambda _1 \left(\frac{b}{r_+}\right){}^{2/\beta }}{q^3 (\beta -2) (\beta -1)}\\
        & +\frac{12 \pi  \beta ^2 \varepsilon  \log (16 \pi ) \lambda _1 \left(\frac{b}{r_+}\right){}^{2/\beta }}{(\beta -2) (\beta +2) r_+}+\frac{36 \pi  \beta  \varepsilon  \log (16 \pi ) \lambda _1 \left(\frac{b}{r_+}\right){}^{2/\beta }}{(\beta -2) (\beta +2) r_+}+\frac{24 \pi  \varepsilon  \log (16 \pi ) \lambda _1 \left(\frac{b}{r_+}\right){}^{2/\beta }}{(\beta -2) (\beta +2) r_+}+\frac{36 \pi  \beta ^2 \varepsilon  \lambda _1 \left(\frac{b}{r_+}\right){}^{2/\beta }}{\left(\beta ^2-4\right) r_+}\\
        & +\frac{84 \pi  \beta  \varepsilon  \lambda _1 \left(\frac{b}{r_+}\right){}^{2/\beta }}{\left(\beta ^2-4\right) r_+}+\frac{48 \pi  \varepsilon  \lambda _1 \left(\frac{b}{r_+}\right){}^{2/\beta }}{\left(\beta ^2-4\right) r_+}+\frac{3 \beta ^2 \varepsilon  \, _2F_1\left(1,1-\frac{2}{3 \beta };2-\frac{2}{3 \beta };\frac{r_+^3}{2 q^3}\right) r_+^3 \lambda _2 \left(\frac{b}{r_+}\right){}^{2/\beta }}{q^6 (\beta -2) (3 \beta -2)}\\
        & +\frac{3 \beta  \varepsilon  \, _2F_1\left(1,1-\frac{2}{3 \beta };2-\frac{2}{3 \beta };\frac{r_+^3}{2 q^3}\right) r_+^3 \lambda _2 \left(\frac{b}{r_+}\right){}^{2/\beta }}{q^6 (\beta -2) (3 \beta -2)}+\frac{6 \varepsilon  \, _2F_1\left(1,1-\frac{2}{3 \beta };2-\frac{2}{3 \beta };\frac{r_+^3}{2 q^3}\right) r_+^3 \lambda _2 \left(\frac{b}{r_+}\right){}^{2/\beta }}{q^6 (\beta -2) (3 \beta -2)}\\& +\frac{18 \beta ^2 \varepsilon  \, _2F_1\left(2,1-\frac{2}{3 \beta };2-\frac{2}{3 \beta };-\frac{r_+^3}{q^3}\right) r_+^3 \lambda _2 \left(\frac{b}{r_+}\right){}^{2/\beta }}{q^6 (\beta -2) (3 \beta -2)}-\frac{3 \beta ^2 \varepsilon  \lambda _2 \left(\frac{b}{r_+}\right){}^{2/\beta }}{q^3 (\beta -2)}-\frac{9 \beta  \varepsilon  \lambda _2 \left(\frac{b}{r_+}\right){}^{2/\beta }}{q^3 (\beta -2)}-\frac{6 \varepsilon  \lambda _2 \left(\frac{b}{r_+}\right){}^{2/\beta }}{q^3 (\beta -2)}\\&-\frac{24 \pi ^2 \varepsilon  r_+ \left(\frac{b}{r_+}\right){}^{2/\beta }}{(\beta -2)^2}-\frac{108 \pi ^2 \beta ^2 \varepsilon  \, _2F_1\left(1,\frac{\beta -2}{3 \beta };\frac{4}{3}-\frac{2}{3 \beta };-\frac{r_+^3}{q^3}\right) r_+ \left(\frac{b}{r_+}\right){}^{2/\beta }}{(\beta -2)^2}\\&-\frac{108 \pi ^2 \beta  \varepsilon  \, _2F_1\left(1,\frac{\beta -2}{3 \beta };\frac{4}{3}-\frac{2}{3 \beta };-\frac{r_+^3}{q^3}\right) r_+ \left(\frac{b}{r_+}\right){}^{2/\beta }}{(\beta -2)^2}\\&-\frac{9 \pi  \beta  \varepsilon  \, _2F_1\left(1,\frac{2 (\beta -1)}{3 \beta };\frac{5}{3}-\frac{2}{3 \beta };\frac{r_+^3}{2 q^3}\right) r_+^2 \lambda _1 \left(\frac{b}{r_+}\right){}^{2/\beta }}{q^3 (\beta -2) (\beta -1)}-\frac{54 \pi  \beta ^2 \varepsilon  \, _2F_1\left(2,\frac{2 (\beta -1)}{3 \beta };\frac{5}{3}-\frac{2}{3 \beta };-\frac{r_+^3}{q^3}\right) r_+^2 \lambda _1 \left(\frac{b}{r_+}\right){}^{2/\beta }}{q^3 (\beta -2) (\beta -1)}\\&-\frac{27 \pi  \beta ^2 \varepsilon  \, _2F_1\left(2,\frac{2 (\beta -1)}{3 \beta };\frac{5}{3}-\frac{2}{3 \beta };-\frac{r_+^3}{q^3}\right) \log (16 \pi ) r_+^2 \lambda _1 \left(\frac{b}{r_+}\right){}^{2/\beta }}{q^3 (\beta -2) (\beta -1)}-\frac{12 \beta  \varepsilon  \, _2F_1\left(1,1-\frac{2}{3 \beta };2-\frac{2}{3 \beta };-\frac{r_+^3}{q^3}\right) r_+^3 \lambda _2 \left(\frac{b}{r_+}\right){}^{2/\beta }}{q^6 (\beta -2) (3 \beta -2)}\\&-\frac{24 \pi ^2 q^3 \varepsilon  \left(\frac{b}{r_+}\right){}^{2/\beta }}{(\beta -2) (\beta +1) r_+^2}-\frac{12 \pi ^2 q^3 \beta ^2 \varepsilon  \left(\frac{b}{r_+}\right){}^{2/\beta }}{(\beta -2) (\beta +1) r_+^2}-\frac{36 \pi ^2 q^3 \beta  \varepsilon  \left(\frac{b}{r_+}\right){}^{2/\beta }}{(\beta -2) (\beta +1) r_+^2}+6 \pi ^2 \Lambda  r_+^3-24 \pi  \Lambda  r_+ \lambda _1+\frac{12 \pi  (\log (16 \pi )+3) \lambda _1}{r_+}\\&-\frac{1}{(\beta -2) r_+ \left(q^3+r_+^3\right)}\left(6 \pi  \log \left(\frac{\left(r_+^3-2 q^3\right) \left(\varepsilon  \left(2 q^3 (\beta +1)-(\beta -2) r_+^3\right) \left(\frac{b}{r_+}\right){}^{2/\beta }+(\beta -2) \left(\Lambda  r_+^5-r_+^3+2 q^3\right)\right){}^2}{(\beta -2)^2 r_+^3 \left(q^3+r_+^3\right){}^2}\right)\right) \\&\left(2 \left(\beta  \varepsilon  \left(\frac{b}{r_+}\right){}^{2/\beta }+\varepsilon  \left(\frac{b}{r_+}\right){}^{2/\beta }+\beta -2\right) q^3+(\beta -2) r_+^3 \left(-\varepsilon  \left(\frac{b}{r_+}\right){}^{2/\beta }+\Lambda  r_+^2-1\right)\right) \lambda _1)\\& +6 \pi  r_+ \left(\Lambda  (\log (16 \pi )+4) \lambda _1+\pi \right)+\frac{6 \log \left(r_+\right) \lambda _2}{q^3}\\& +\frac{2 \left(9 \pi ^2 \left(\Lambda  q^6+3 r_+ q^3\right)-3 \pi  (\log (16 \pi )+2) r_+ \left(\Lambda  q^3+3 r_+\right) \lambda _1+\left(\Lambda  r_+^2-3\right) \lambda _2\right)}{q^3+r_+^3}\\& +\frac{4 \Lambda  r_+ \left(3 \pi  \left(r_+^3-2 q^3\right) \lambda _1 q^3+r_+ \left(q^3+r_+^3\right) \lambda _2\right)}{-2 q^6-r_+^3 q^3+r_+^6}-\frac{2 \log \left(r_+^3-2 q^3\right) \lambda _2}{q^3}-\frac{12 \pi ^2 q^3}{r_+^2}\Bigg].
        \label{Fc_explicit}
\end{aligned}
\end{equation}
\end{itemize}
\begin{itemize}
\item {\bf Specific heat:}
Uncorrected specific heat is given by
\begin{equation}
    C_0 = \frac{2 \pi  \beta  \left(q^3+r_+^3\right){}^2 \left(\varepsilon  \left(\frac{b}{r_+}\right){}^{2/\beta } \left((\beta -2) r_+^3-2 (\beta +1) q^3\right)-(\beta -2) \left(2 q^3+\Lambda  r_+^5-r_+^3\right)\right)}{\varepsilon  r_+ \left(\frac{b}{r_+}\right){}^{2/\beta } C_{01}-(\beta -2) \beta  r_+ \left(-2 q^6+2 q^3 r_+^3 \left(2 \Lambda  r_+^2-5\right)+\Lambda  r_+^8+r_+^6\right)}.
\end{equation}
where, $C_{01}=\left(2 (\beta +1) (\beta +2) q^6+2 (\beta  (5 \beta +3)+4) q^3 r_+^3-\left(\beta ^2-4\right) r_+^6\right)$
\item {\bf Gibbs free energy:}
The uncorrected Gibbs free energy is given by
\begin{equation}
    G = M -T_H S.
\end{equation}
For this case, it becomes:
\begin{equation}
    G = \frac{3 \varepsilon  \left(\frac{b}{r_+}\right){}^{2/\beta } G_{1}+(\beta -2) \left(-2 q^6 \left(\Lambda  r_+^2+3\right)+2 q^3 r_+^3 \left(12-5 \Lambda  r_+^2\right)+r_+^6 \left(\Lambda  r_+^2+3\right)\right)}{12 (\beta -2) r_+^2 \left(q^3+r_+^3\right)}.
    \label{G_unc}
\end{equation}
where, $G_{1}\left(-2 (\beta +2) q^6+2 (4 \beta -1) q^3 r_+^3+(\beta +2) r_+^6\right)$
\end{itemize}
\section*{Acknowledgments}
SB is grateful to Tezpur University for providing financial support through the Institutional Fellowship. She also shows her sincere gratitude towards her colleagues from Astrophysical Plasma and Nonlinear Dynamics Research Laboratory (APNDRL). DJG acknowledges the contribution of the COST Action CA21136  -- ``Addressing observational tensions in cosmology with systematics and fundamental physics (CosmoVerse)". IUCAA Associateship to PKK is thankfully acknowledged.

\section*{Credit Authorship Contribution statement}
{\bf Shyamalee Bora}: Calculations, Formal analysis, Investigation, Writing- original draft. {\bf Dhruba Jyoti Gogoi}: Conceptualisation, Investigation, Discussion, Methodology, Final editing. {\bf Pralay Kumar Karmakar}: Project administration, Supervision, Physical interpretation, Corresponding author, Proof reading, Final editing.

\section*{Declaration of competing interest}
The authors declare that they have no known competing financial interests or personal relationships that could have appeared to influence the work reported in this manuscript.

\section*{Data Availability Statement}
There are no new data associated with this article.

\bibliography{references}
\end{document}